\documentclass[a4paper,11pt]{article}
\pdfoutput=1 

\usepackage{jheppub} 
\usepackage[T1]{fontenc} 
\usepackage{here}

\title{\boldmath Observation of Reactor Antineutrino Disappearance Using Delayed Neutron Capture on Hydrogen at RENO}

\collaboration{The RENO Collaboration}

\author[a]{C. D. Shin,}
\author[a]{Zohaib Atif,}
\author[a]{G. Bak,}
\author[b]{J. H. Choi,}
\author[c]{H. I. Jang,}
\author[d]{J. S. Jang,}
\author[e]{S. H. Jeon,}
\author[a,1]{K. K. Joo,\note{Corresponding author.}}
\author[g]{K. Ju,}
\author[e]{D. E. Jung,}
\author[e]{J. G. Kim,}
\author[a]{J. Y. Kim,}
\author[f,1]{S. B. Kim,}
\author[f]{S. Y. Kim,}
\author[i]{W. Kim,}
\author[f]{E. Kwon,}
\author[f]{D. H. Lee,}
\author[f]{H. G. Lee,}
\author[f]{Y. C. Lee,}
\author[a]{I. T. Lim,}
\author[a]{D. H. Moon,}
\author[b]{M. Y. Pac,}
\author[e]{C. Rott,}
\author[f]{H. Seo,}
\author[a]{J. H. Seo,}
\author[e]{J. W. Seo,}
\author[f,2]{S. H. Seo,\note{Now at Institute for Basic Science, Yuseong-gu, Daejeon, 34126, Korea}}
\author[h]{B. S. Yang}
\author[f]{J. Y. Yang,}
\author[g,h]{J. Yoo}
\author[e]{and I. Yu}

\affiliation[a]{Institute for Universe and Elementary Particles, Department of Physics, Chonnam National University, Gwangju 61186, Korea          }
\affiliation[b]{Institute for High Energy Physics, Dongshin University, Naju 58245, Korea                     }
\affiliation[c]{Department of Fire Safety, Seoyeong University, Gwangju 61268, Korea              }
\affiliation[d]{GIST College, Gwangju Institute of Science and Technology, Gwangju 61005, Korea         }
\affiliation[e]{Department of Physics, Sungkyunkwan University, Suwon 16419, Korea                }
\affiliation[f]{Department of Physics and Astronomy, Seoul National University, Seoul 08826, Korea }
\affiliation[g]{Department of Physics, KAIST, Daejeon 34141, Korea}
\affiliation[h]{Institute for Basic Science, Daejeon 34047, Korea }
\affiliation[i]{Department of Physics, Kyungpook National University, Daegu 41566, Korea  }

\emailAdd{kkjoo@chonnam.ac.kr, sbk@snu.ac.kr}

\abstract{The Reactor Experiment for Neutrino Oscillation (RENO) experiment has been taking data using two identical liquid scintillator detectors of 44.5 tons since August 2011. The experiment has observed the disappearance of reactor neutrinos in their interactions with free protons, followed by neutron capture on hydrogen. Based on 1500 live days of data taken with 16.8 GW$_{th}$ reactors at the Hanbit Nuclear Power Plant in Korea, the near (far) detector observes 567690 (90747) electron antineutrino candidate events with a delayed neutron capture on hydrogen. 
This provides an independent measurement of $\theta_{13}$ and a consistency check on the validity of the result from n-Gd data. Furthermore, it provides an important cross-check on the systematic uncertainties of the n-Gd measurement.
Based on a rate-only analysis, we obtain sin$^{2}$2$\theta _{13}$= 0.087 $\pm$ 0.008 (stat.) $\pm$ 0.014 (syst.).}


\begin{document} 
\maketitle
\flushbottom

\section{Introduction}
\label{sec:intro}

In the framework of three flavors, neutrino oscillation is described by the Pontecorvo-Maki-Nakagawa-Sakata matrix with three mixing angles ($\theta_{12}$, $\theta_{23}$, and $\theta_{13}$), two mass-squared differences, and one CP phase angle \cite{CKM1,CKM2}.
The smallest mixing angle $\theta _{13}$ is unambiguously determined from the reactor electron antineutrino  ($\overline{\nu}_e$) disappearance, observed by three reactor experiments using Gadolinium (Gd)-loaded liquid scintillator (LS) \cite{RENOPRL1,DCPRL1,DBPRL1}. 
The successful measurement of $\theta _{13}$ serves as the very first step to the complete understanding of the fundamental nature and implications of neutrino masses and mixing parameters. A rather large value $\theta _{13}$ opens an exciting opportunity to search for CP violation in the lepton sector and to determine the neutrino mass hierarchy \cite{Masshie}. 
A precise measurement of $\theta _{13}$ by a reactor experiment would provide important insight into the determination of the leptonic CP violating phase 
if the accelerator beam results are combined \cite{T2K1,NOVA1}.

RENO is the first reactor experiment to take data with two identical near and far detectors in operation from August 2011.
The RENO's earlier measurements of $\theta _{13}$ \cite{RENOPRL1,RENOPRL2,RENOPRD} are based on detecting reactor $\overline{\nu}_e$ through the inverse beta decay (IBD) interaction, $\overline{\nu}_e + p \rightarrow e^+  + n$, with a $\sim$26 $\mu$s-delayed signal of $\sim$8 MeV $\gamma$-rays from neutron capture on Gd (n-Gd). 
The delayed coincidence with a prompt positron signal significantly reduces the background events coming from natural radioactivity predominantly below 3 MeV \cite{NatRad}.

RENO is also sensitive to detecting reactor $\overline{\nu}_e$  by the coincidence of a prompt positron signal and a $\sim$200 $\mu$s-delayed $\gamma$-ray of 2.2 MeV from neutron capture on hydrogen (n-H).
Clear detection of the n-H delayed signal is possible due to the successful purification of LS and detector materials, use of the lower radioactive photomultiplier tube (PMT) glass, and effective shielding against $\gamma$-rays from the surrounding rocks. 
Furthermore, because of the better understanding of various backgrounds in the n-H data sample, the systematic uncertainty is sufficiently reduced to determine the value of $\theta_{13}$. More than twice of IBD n-H events compared to IBD n-Gd events are produced in the RENO detector.

The number of protons is estimated to be (1.189 $\pm$ 0.008)$\times 10^{30}$ in the Gd-loaded LS and (2.110 $\pm$ 0.015)$\times 10^{30}$ in the Gd-unloaded LS where uncertainties on density meter and solvent composition are both included, based on the measured scintillator hydrogen fraction \cite{proton}. 
By employing the n-H detection method, we can use $\sim$2.8 times more target than that used by the n-Gd measurement. 
This corresponds to $\sim$2.3 more production of IBD n-H events.

RENO has performed a measurement of $\theta_{13}$ using a significantly large IBD n-H data set.  
The motivation of the n-H analysis is to provide an independent measurement of $\theta _{13}$ and to check the consistency with the n-Gd measurement. 
Moreover, it provides a valuable cross-check on the systematic uncertainties of the n-Gd $\theta _{13}$ measurement.
Recently, RENO has published papers \cite{RENOPRL3} on the improved measurement of $\theta_{13}$ from IBD n-Gd analysis using $\sim$2200 live days of data.
Daya Bay and Double Chooz Collaborations reported their first $\theta_{13}$ measurements using IBD n-H events \cite{DBnH, DCnH}.
 
In this paper, we present the RENO's first measured values of $\theta_{13}$ from the reactor $\overline{\nu}_e$ disappearance observed in the IBD interactions with neutron capture on hydrogen in the near and far detectors based on $\sim$1500 live days of data taken from 11 August 2011 to 23 April 2017. 


\section{The RENO experiment}

\begin{figure}[tbp]
	\centering
	\includegraphics[width=0.75\textwidth]{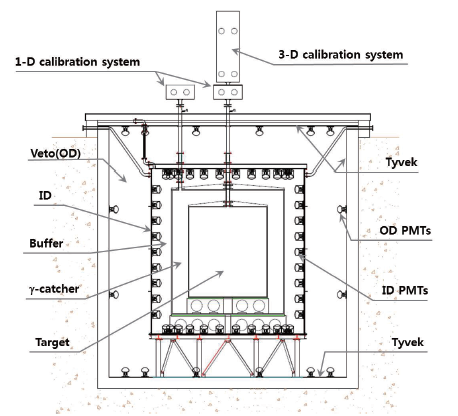}

	\caption{
A schematic view of the RENO detector consisting of four concentric cylindrical components, i.e., target, $\gamma$-catcher and buffer for ID and veto for OD.	
The ID is contained in a cylindrical stainless vessel of 5.4 m in diameter and 5.8 m in height, and the OD is surrounded by a cylindrical concrete
    of 8.8 m in height and 8.4 m in diameter. 
	The diameter of the whole detector is 8.4 m and the height is 8.8 m. There are 354 (67) 10-inch PMTs in ID (OD).}
	\label{fig:detector}
\end{figure}

Six pressurized water reactors at Hanbit Nuclear Power Plant in South Korea, each with a maximum thermal output of 2.8 GW$_{\rm th}$, are situated in a linear array spanning 1.3 km with equal spacings.
Identical near and far antineutrino detectors in the RENO experiment are located at 294 m and 1383 m, respectively, from the center of six reactor cores of the Hanbit reactor, providing the maximum thermal output of 16.8 GW$_{\rm th}$. 
The reactor-flux weighted baseline is 408.56 m for the near detector, and 1443.99 m for the far detector. 
The near (far) detector is under 120 (450) meters of water-equivalent rock overburden. 
Through use of an identical design for both detectors, a number of systematic uncertainties associated with the measurement of $\theta_{13}$ cancel each other out in the far-to-near ratio measurement.

Each RENO detector consists of a main inner detector (ID) and an outer veto detector (OD). 
From the innermost to the outermost, the three detector components of target, $\gamma$-catcher and buffer belong to ID as shown in Fig. \ref{fig:detector}.
The liquid scintillator (LS) is produced as a mixture of linear alkyl benzene (LAB, C$_{n}$H$_{2n+1}-$C$_6$H$_5$, n = 10$\sim$13) with fluors of 3 g/$\textit{l}$ of 2,5-diphenyloxazole (PPO) and 30 mg/$\textit{l}$ of 1,4-bis(2-methylstyryl)benzene (bis-MSB). A Gd-carboxylate compound using 3,5,5-trimethylhexanoic acid (TMHA) was developed for the best Gd loading efficiency into LS and its long term stability \cite{GdLS1, GdLS2}. The main detector is contained in a cylindrical stainless steel vessel that houses two nested cylindrical acrylic vessels \cite{RENODet}. The innermost acrylic vessel holds the 18.7 m$^3$ (16.0 tons) $\sim$0.1\% Gd-loaded LS as a neutrino target. It is surrounded by a $\gamma$-catcher (GC) region with a 60 cm thick layer of Gd-unloaded LS with a volume of 33.2 m$^3$ (29.0 tons) inside an outer acrylic vessel. Outside of the $\gamma$-catcher, there is a 70 cm thick buffer region filled with 62.7 tons of mineral oil. The light signals emitted from the particles are detected by 354 low background 10-inch PMTs \cite{RENOPMT} mounted on the inner wall of the stainless-steel container. The 1.5 m thick OD region is filled with highly purified water and equipped with 67 10-inch PMTs mounted on the wall of the concrete veto vessel so as to catch the water Cherenkov light.

\section{Reconstruction}
\subsection{Vertex reconstruction}
\begin{figure}[tbp]
 	\centering
 	\includegraphics[width=0.85\textwidth]{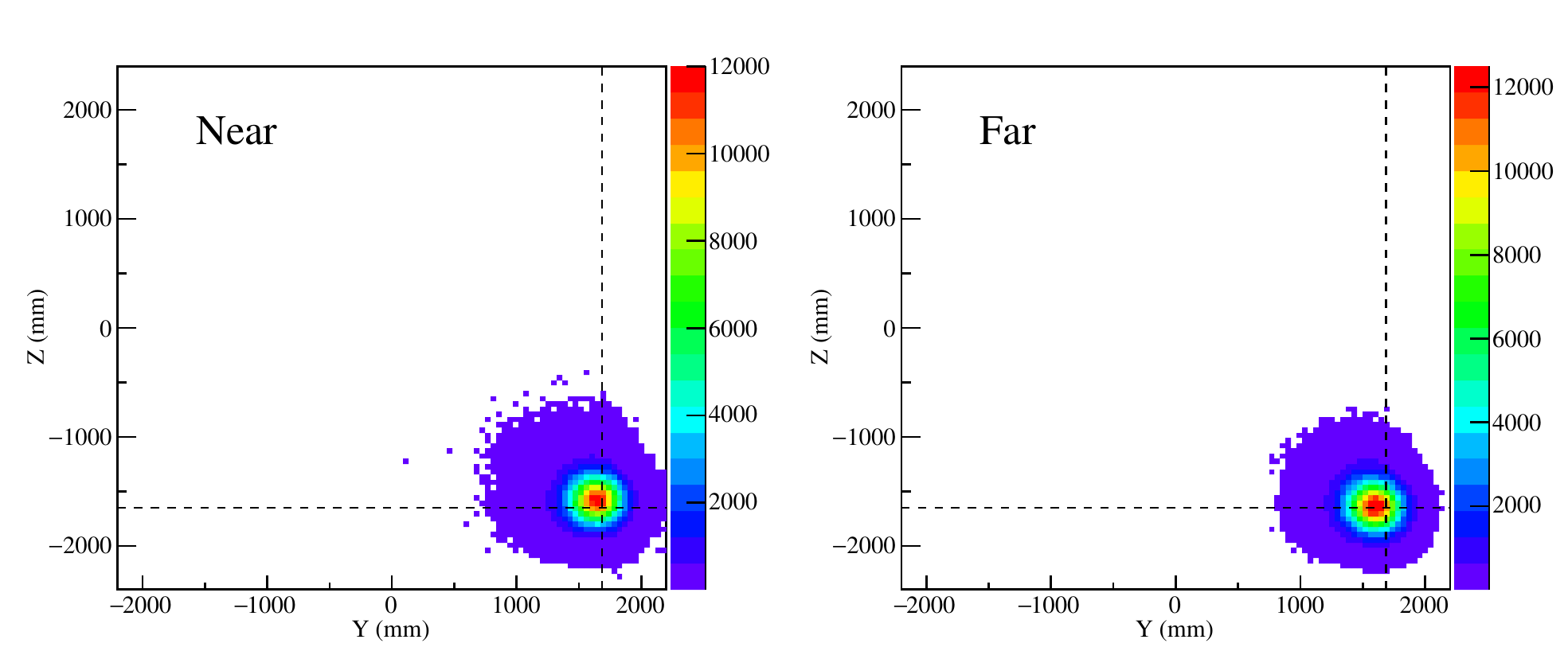}	
 	\caption{
 		Reconstructed vertices of $^{60}$Co source in GC. The RENO uses Cartesian coordinates of (x, y, z), and the detector center is (0,0,0). The point crossed by black dashed lines is actual position at the source location of GC in detector.
 	}
 	\label{fig:vert_source}
 \end{figure}

 The event vertex information is used to distinguish the target and GC signals. In addition, the difference in distance between prompt and delayed candidates is useful for eliminating and measuring accidental backgrounds. 
 The event vertex is reconstructed using the observed charge information of individual PMT. 
 A basic position reconstruction algorithm is defined, using a charge centroid calculation. This method has been used for reconstruction in many existing detectors, typically as a seed for a more sophisticated algorithm 
 in an isotopic light of a scintillator detector. 
 In order to reconstruct event vertex, position of all the hit PMTs is calculated as a charge weighted average.
 
 \begin{eqnarray}
 \vec{r}_{vtx} = \dfrac{\Sigma_{i}(Q_{i}\cdot\vec{r}_{i})}{\Sigma_{i}(Q_{i})},
 \end{eqnarray}
 where $\vec{r}_{vtx}$ is a reconstructed vertex of each event, $i$ is an index of each PMT, $Q_{i}$ is the charge collected by the $i$-th PMT, and $\vec{r}_{i}$ is a position vector of the PMT. This method works well in spherical, fully symmetric detectors. In the RENO detector, there is a difference between true and reconstructed vertex due to the geometrical effects as a cylindrical structure. A correction factor is calculated using a Monte Carlo calculation that includes geometrical shape of detector and the attenuation length of materials. The reconstructed vertex is confirmed through source data and match well with the actual location of source data as shown in Fig. \ref{fig:vert_source}. These results depend on the energy of the source because it uses charge information. 
 The vertex resolution is less than $\sim$17 cm at the 1 MeV, and improves at higher energies.

\subsection{Energy reconstruction}

\begin{figure}[tbp]
	\centering
	\includegraphics[width=0.75\textwidth]{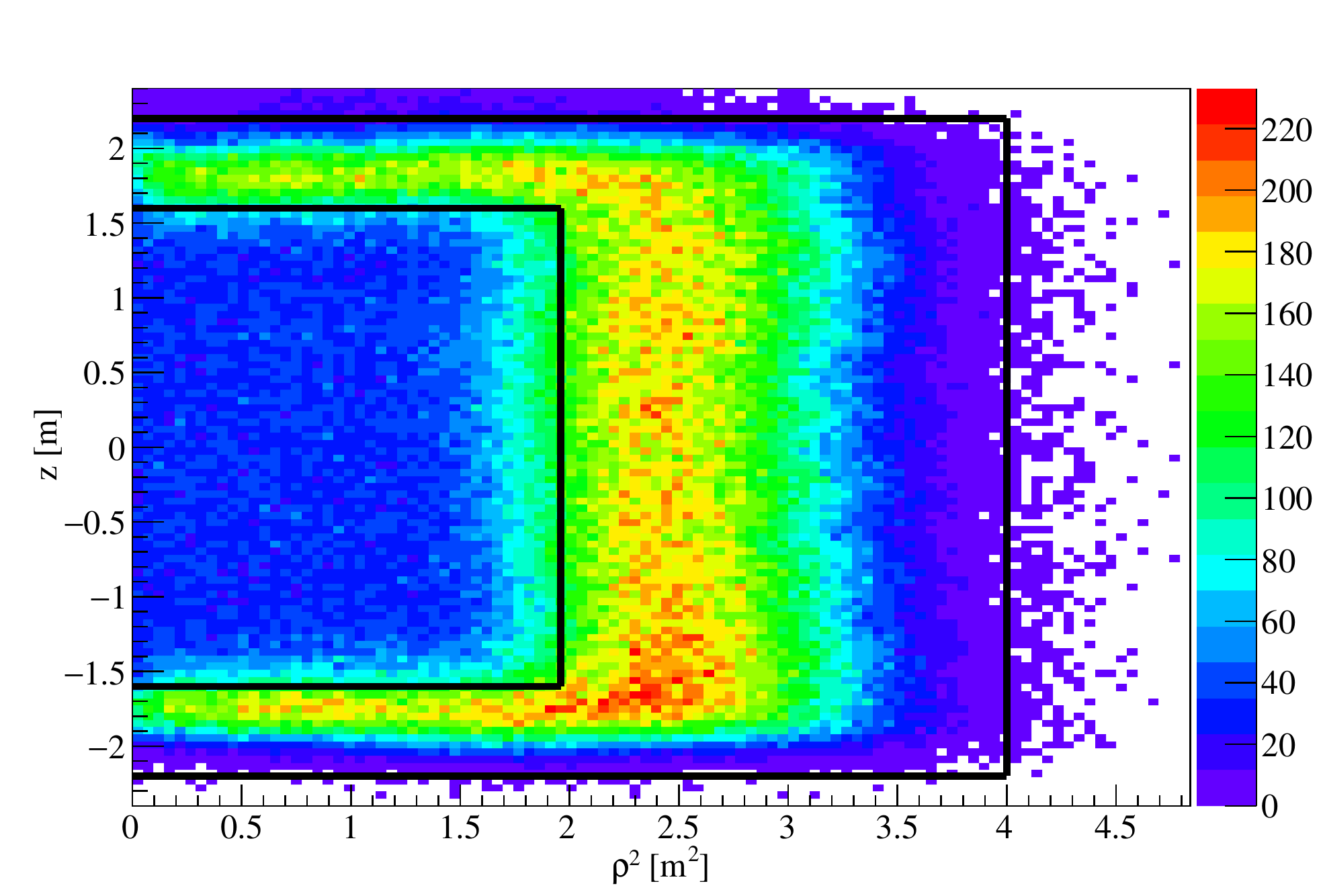}	
	\caption{
		Vertex distribution of n-H IBD events in the height ($z$) and radius ($\rho$) plane. 
		The number of reconstructed IBD events are represented by color. The inner (outer) solid line represents the target (GC) boundary. Neutrons are mostly captured by Gd in the target where the n-H delayed candidates are much fewer than in the GC. The buffer surrounding the GC is filled with non-scintillating mineral oil and expects no events in the region.
			}
		

	\label{fig:vertex}
\end{figure}
\begin{figure}[t]
	\centering
	\includegraphics[width=0.80\textwidth]{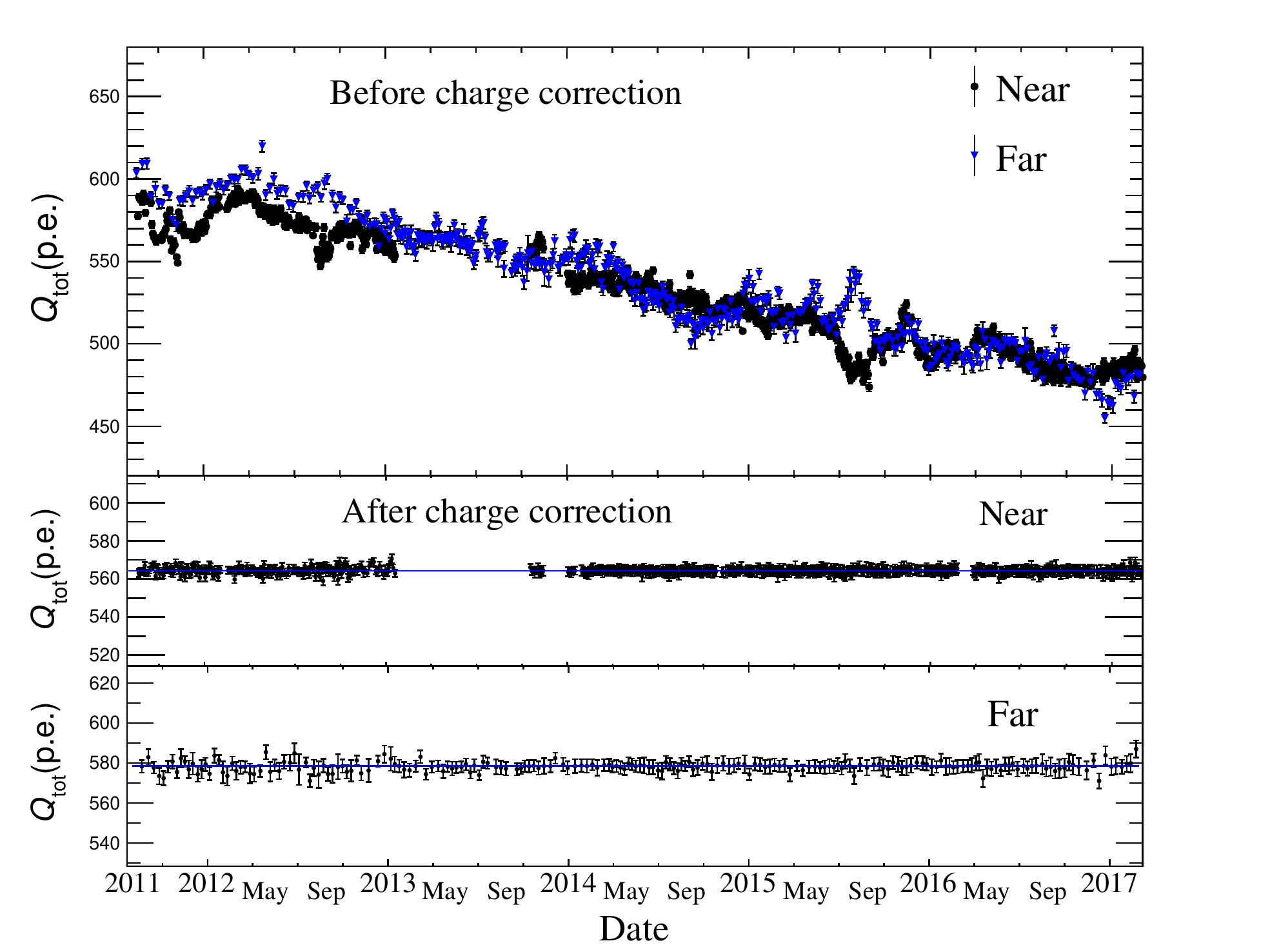}
	
	\caption{Raw charge variation (top panel) and corrected charge stability (bottom panel) of n-H delayed signal in the GC. The raw $Q_{\rm tot}$ decreases over time and the corrected one becomes close to the reference value in the initial period. Roughly 400 days of near detector data are unused for this analysis because of an electrical noise coming from an uninterruptible power supply (UPS).
}
	\label{fig:s2_stab}
\end{figure}

An event is observed by collecting scintillation lights in the PMTs. An electronic board with ethernet (QBEE) based on a charge-to-time converter (QTC) takes a PMT analog signal and converts it to a digital value. An analog-to-digital converter (ADC) value is changed to a value in pC. The charge of a PMT is converted to that in photoelectron (p.e.). A measured pC-to-p.e. conversion factor is obtained from the radioactive source of $^{137}$Cs that delivers roughly a single p.e. to a PMT, corresponding to $\sim$1.6 pC. The energy of an event is determined by a total charge ($Q_{tot}$) that is the sum of charges in hit PMTs with more than 0.3 p.e. in a time window of –100 to 50 ns. 
Neutrons from IBD events are captured only by hydrogen in the GC while roughly 15\% of neutrons are captured by hydrogen in the target. Thus most of the n-H IBD events occur in the GC region as shown in Fig. \ref{fig:vertex}.

The energy calibration is performed separately for the target and GC
regions because of their different optical properties. The raw $Q_{\rm tot}$ of n-H delayed
shows a gradual decrease in time as shown in Fig. \ref{fig:s2_stab}. The observed $Q_{\rm tot}$ is
reduced by $\sim$15\% at most of the initial operation value. This is caused by removing
the malfunctioning or flashing PMTs and the decrease of the LS
attenuation length \cite{RENOatt}. The attenuation length decrease is due to loose air
tightening around the detector chimney region where oxygen or moisture are
introduced into the detector. The attenuation length no longer decreases after
careful air-shielding with nitrogen gas.
%
The raw charge variation is corrected over time by using the peak value of the delayed signal energy. This should be the same as the reference value measured at the initial period of the experiment. After the charge correction, the bottom plot of Fig. \ref{fig:s2_stab} shows corrected $Q_{\rm tot}$ of the n-H delayed signal becomes recovered to the reference value.

\subsection{Muon energy estimation} 
 Cosmogenic muons produce backgrounds even in the underground detector. The intrinsic muon energy can not be measured accurately because they pass through the detector often without depositing the entire energy. However, a cosmic muon deposits energy proportional to its path length. The muon deposit energy ($E_{\mu}$) is reconstructed from the observed $Q_{\rm tot}$ using a conversion factor of 250 p.e. per MeV. The muon deposit energy can be measured up to $\sim$1.7 GeV, due to the saturation of DAQ electronics in RENO. The minimum deposit energy of muon identified is 70 MeV. The muon rate is measured to be 117.5 Hz (13.1 Hz) for the near (far) detector.

\section{Energy calibration}

\begin{figure}[tbp]
	\centering
	\includegraphics[width=0.65\textwidth]{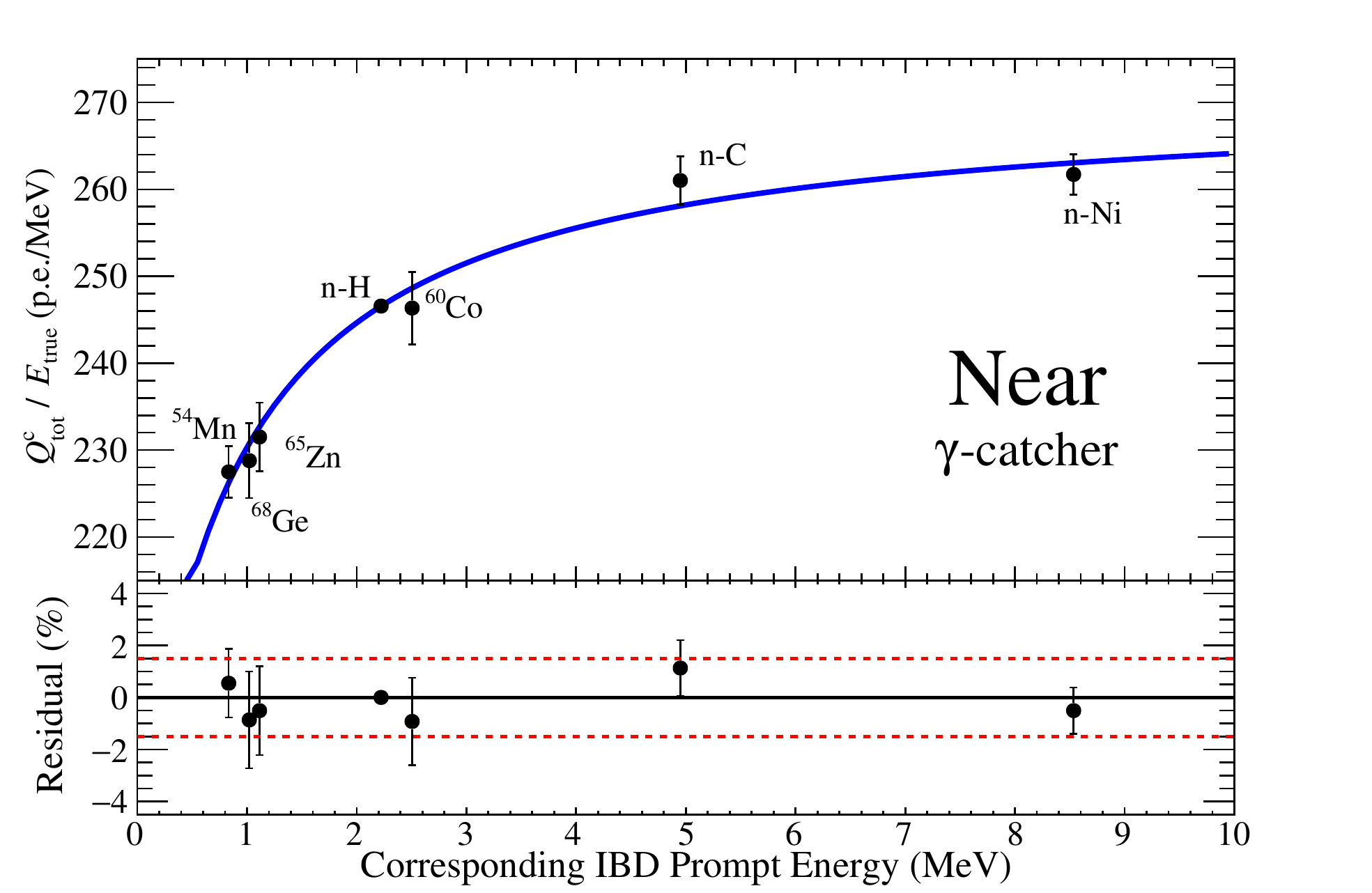}
	\includegraphics[width=0.65\textwidth]{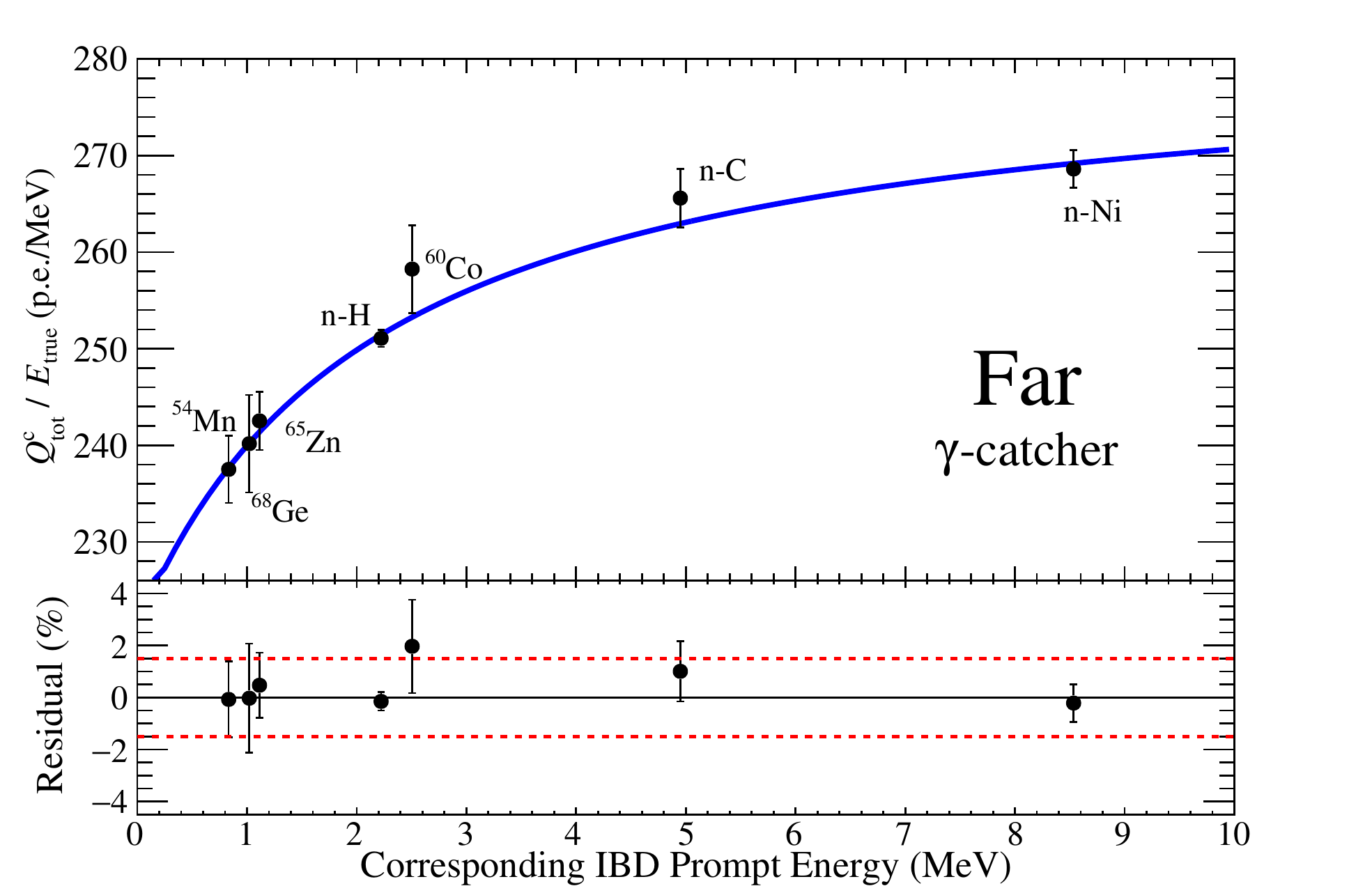}
	
	\caption{Energy conversion for the IBD prompt signal observed in the GC regions of the near and far detectors. It is obtained from the visible energies of $\gamma$-rays coming from various radioactive sources and n-H delayed signal. The curves are the best fits to the data points using a non-linear function. The lower panels show the fractional residuals of calibration data points with respect to the best fit.}
	\label{fig:conv}
\end{figure}
\begin{figure}[tbp]
	\centering
	\includegraphics[width=0.75\textwidth]{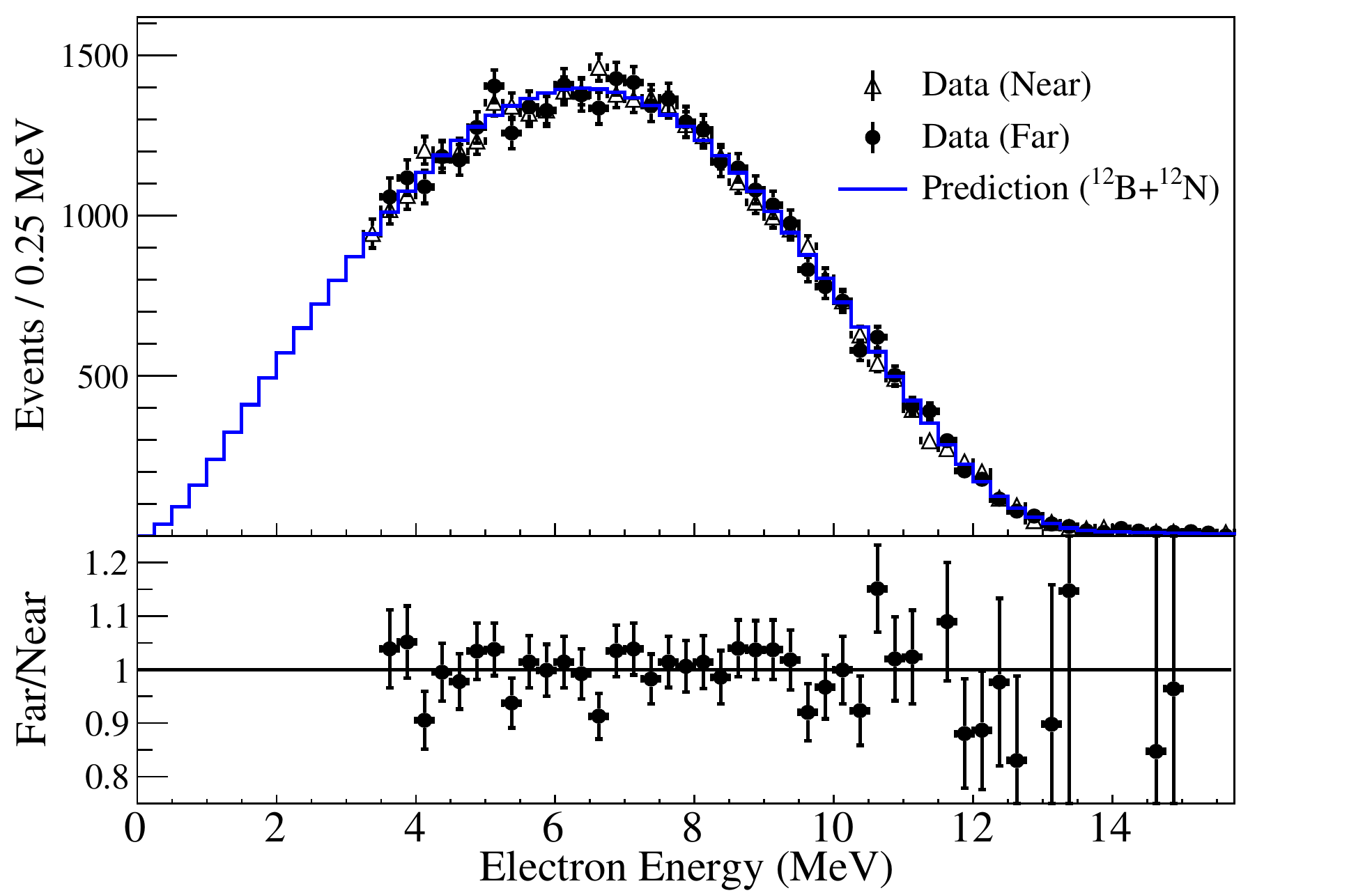}
	
	\caption{Comparison of data and simulated energy spectra of the $\beta$-decay electrons from unstable $^{12}$B and $^{12}$N isotopes produced by cosmic muons. The measured spectra in the near and far detectors are also compared after normalization to the total number of events in the far detector. The far-to-near ratio of the spectra is shown in the lower panel. The good agreement demonstrates identical performance of the near and far detectors and the MC reproducibility in the energy reconstruction.}
	\label{fig:BNconv}
\end{figure}

The energy calibration for the target region is accomplished using several radioactive sources and the IBD n-H and n-Gd events and well described in Ref. \cite{RENOPRL2, RENOPRD}. The energy calibration for the GC region is made using radioactive sources of $^{54}$Mn, $^{68}$Ge, $^{65}$Zn, $^{60}$Co, $^{241}$Am-Be and $^{252}$Cf-Ni, and n-H from IBD signals. 
%
Various radioactive sources are periodically deployed into the target and GC using a motorized pulley system in the glove box, as shown in Fig. \ref{fig:detector}.	
%
The absolute energy scale of a prompt event is determined using a charge-to-energy conversion function obtained from various radioactive sources described above and neutron capture samples. The observed charges of the source data, taken at a certain detector position, are also corrected for a different charge response of uniformly distributed prompt events. The observed $Q_{\textrm{tot}}$ in the $\gamma$-ray source is different from that of positron with the same kinetic energy. The GLG4SIM Monte Carlo (MC) simulation \cite{GLG4SIM} is used to estimate the difference in the observed $Q_{\textrm{tot}}$ between positron and $\gamma$-ray. Using the difference, the observed $Q_{\textrm{tot}}$ of a $\gamma$-ray is converted to a corresponding $Q_{\textrm{tot}}$ of positron ($Q_{\textrm{tot}}^{e^+}$). Fig. \ref{fig:conv} shows the non-linear response of scintillating energy for the IBD prompt signal that is obtained from various radioactive sources in the GC of near and far detectors. This is mainly due to the quenching effect in the scintillator as well as the Cherenkov radiation. 
The non-linear response is well described by a fitted parametrization and consistent with the MC prediction.
The RENO MC includes various measured optical properties of LS and quenching effect of $\gamma$-ray at low energies.  
The empirical energy conversion function is parameterized as follows to reflect non-linearity due to quenching effect especially in the low energy region.

\begin{eqnarray}
Q_{\textrm{tot}}^{e^+}/E_{\textrm{true}} = a + b/[1 - exp (-cE_{\textrm{true}} + d)],
\end{eqnarray}
\\ 

where $E_{\textrm{true}}$ is the true energy of the prompt signal in MeV, the total kinetic and pair-annihilation energy of positron. The parameters of $a$, $b$, $c$, and $d$ are determined by a fit to the calibration data. According to the energy calibration, the observed charge $Q_{\textrm{tot}}$ is $\sim$230 p.e. per MeV at 1 MeV. 
The fitted parameters of the energy conversion function are listed in Table \ref{teconv}. 
The deviations of all calibration data points with respect to the best fit parametrization are within 2$\%$ as shown in Fig. \ref{fig:conv}.

\begin{table}[tbp]
	\begin{center}
		\begin{tabular*}{0.75\textwidth}{@{\extracolsep{\fill}} c  c  c}\hline\hline
			Parameter& Near & Far\\\hline
			$a$ & 271.1 $\pm$ 2.4 &  281.7 $\pm$ 4.6\\
			$b$ & (1.97 $\pm$ 0.29)$\times$10$^{-2}$ & (2.24 $\pm$ 0.72)$\times$10$^{-2}$\\
			$c$ & (2.62 $\pm$ 0.48)$\times$10$^{-4}$ & (1.67 $\pm$ 0.73)$\times$10$^{-4}$\\
			$d$ & (2.20 $\pm$ 0.53)$\times$10$^{-4}$ & (3.70 $\pm$ 1.25)$\times$10$^{-4}$\\
			\hline\hline
		\end{tabular*}
		\caption{Fitted parameter values of energy conversion functions at the near and far detectors.}
		\label{teconv}
	\end{center}
\end{table}

The $\beta$-decays of cosmogenic $^{12}$B and $^{12}$N isotopes are used to check the validity of the charge-to-energy conversion functions. They are produced by cosmic-muon interaction with carbon in liquid scintillator to emit electrons through $\beta$-decay. Figure \ref{fig:BNconv} shows good agreement of the $\beta$ spectrum between data and MC simulation. This indicates that the energy conversion function works well for the prompt energy reconstruction.

\section{Event selections}

Event selection criteria similar to those of the n-Gd analysis \cite{RENOPRL2,RENOPRD,RENOPRL3} are applied to the n-H IBD candidates. More improved and optimized selection requirements are necessary for reducing a larger background of the n-H delayed signal due to its longer capture time and lower energy than the n-Gd delayed signal. Table \ref{tab:IBDSel} compares the selection criteria between the n-H and n-Gd analyses.
 
\begin{figure}[tbp]
	\centering
	\includegraphics[width=0.65\textwidth]{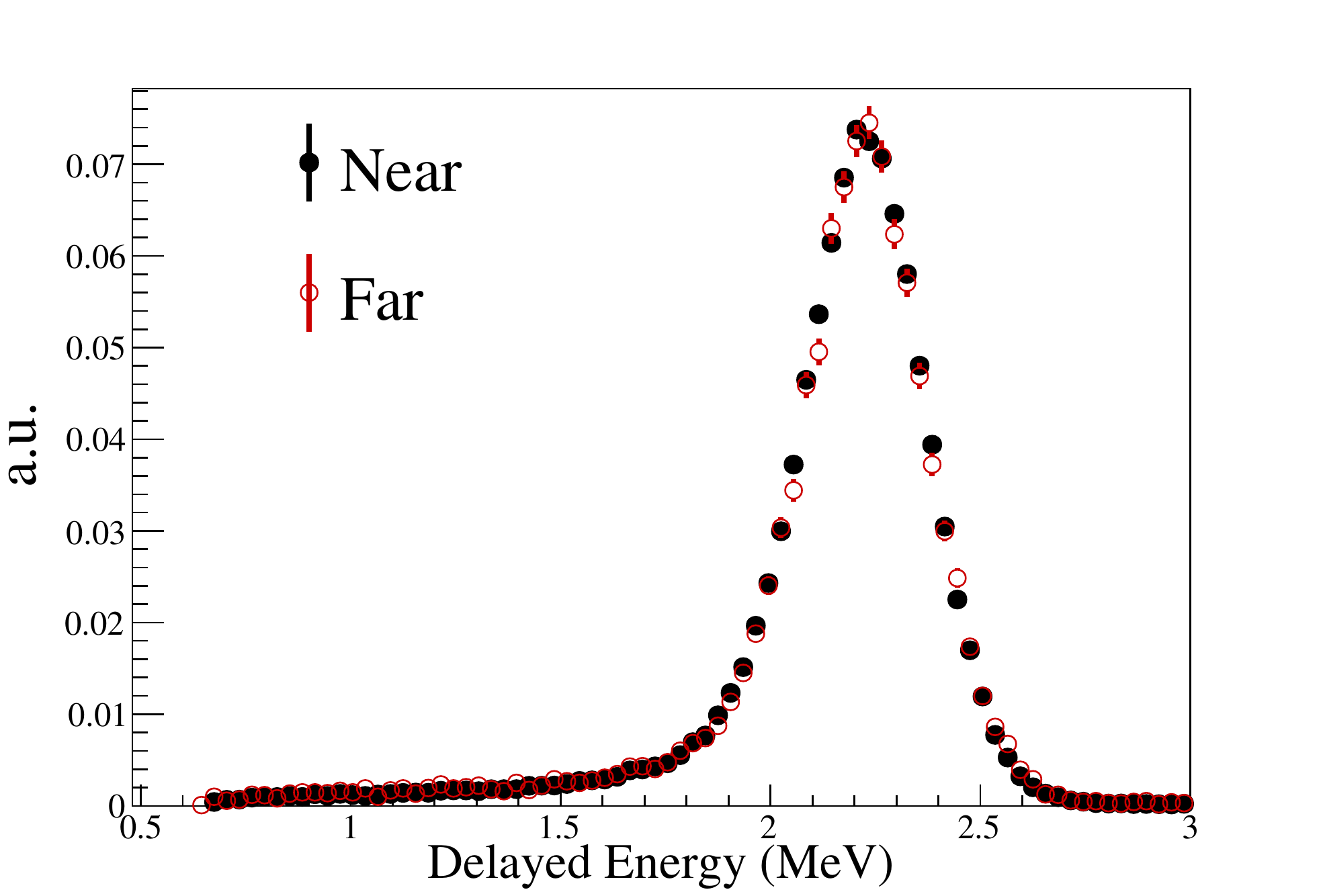}
	\caption{Energy spectrum of delayed signal from neutron capture on hydrogen.}
	\label{fig:nHs2}
\end{figure}

\begin{figure}[tbp]
	\centering
	\includegraphics[width=0.65\textwidth]{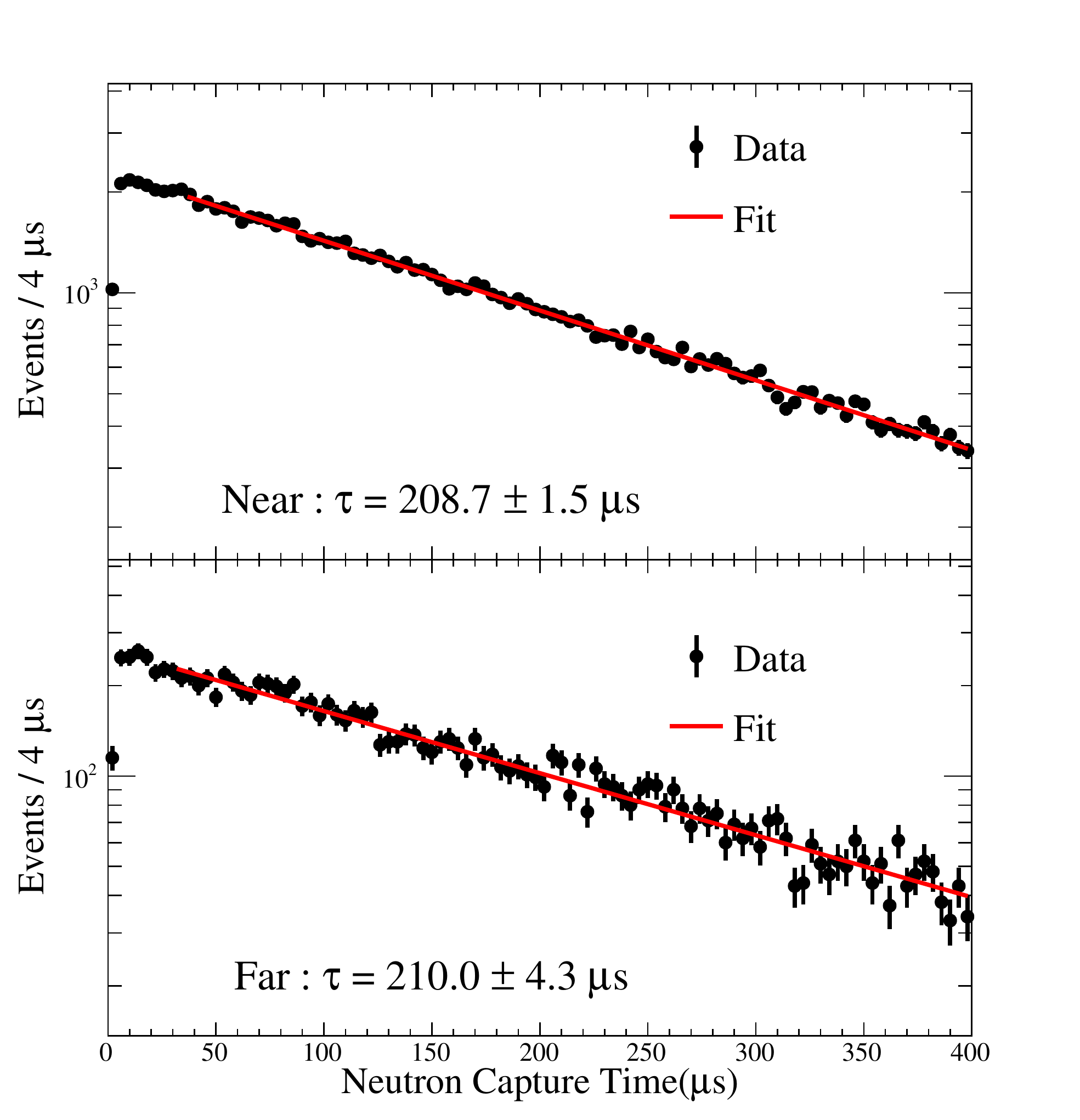}
	\caption{Measured time distributions of neutron capture on hydrogen. The mean capture time is $\sim$200 $\mu$s from the best fit and consistent between the near and far detectors within uncertainties.}
	\label{fig:CT}
\end{figure}

An event trigger is applied using the number of hit PMTs in the buffer region ($N_{\textrm{hit}}$). $N_{\textrm{hit}}$ of selected events is required to be larger than 90 within 50 ns. For removing backgrounds, various criteria are applied as follows: (i) removing events within a 700 ms (500 ms, 200 ms) window following a cosmic muon of $E_{\mu}$ > 1.5 GeV (1.2$\sim$1.5 GeV, 1.0$\sim$1.2 GeV) for the far detector or within a 700 ms (400 ms, 200 ms) window following a cosmic muon of $E_{\mu}$ > 1.6 GeV (1.5$\sim$1.6 GeV, 1.4$\sim$1.5 GeV) for the near detector, or within a 1 ms window following a cosmic muon of $E_{\mu}$ > 70 MeV, or of 20 < $E_{\mu}$ < 70 MeV for $N_{\textrm{hit}}^{\textrm{OD}}$ > 50; (ii) $Q_{\textrm{max}}^{\textrm{prompt}}$/$Q_{\textrm{tot}}^{\textrm{prompt}}$ < 0.07 and $Q_{\textrm{max}}^{\textrm{delayed}}$/$Q_{\textrm{tot}}^{\textrm{delayed}}$ < 0.06, where $Q_{\textrm{max}}$ is the maximum charge of any single ID PMTs; (iii) 0.7 < $E_{\textrm{p}}$ < 12.0 MeV, where $E_{\textrm{p}}$ is the energy of the prompt signal; (iv) 2.223$-$2$\sigma _{\textrm{d}}$ < $E_{\textrm{d}}$ < 2.223$+$2$\sigma _{\textrm{d}}$ MeV, where $E_{\textrm{d}}$ is the energy of delayed candidate, $\sigma _{\textrm{d}}$ is a standard deviation from the delayed energy peak. Figure \ref{fig:nHs2} shows an energy spectrum of clean delayed candidates of $\sim$2.2 MeV $\gamma$-rays from neutron captures on H; 
(v) 2 < $\Delta t$ < 400 $\mu$s, where $\Delta t$ is the time difference between the prompt and delayed candidates. 
Figure \ref{fig:CT} shows the $\Delta t$ distribution of n-H IBD candidates. The best fit value is 208.7$\pm$1.5 (210$\pm$4.3) $\mu$s for the near (far) detector; 
(vi) $\Delta R$ < 450 mm, where $\Delta R$ is the vertex difference between the prompt and delayed candidates; (vii) a timing veto requirement for rejecting coincidence pairs (a) if they are accompanied by any preceding ID or OD trigger within a 500 $\mu$s window before their prompt candidate, (b) if they are accompanied by only the ID or the ID \& OD trigger within a 500$\sim$600 $\mu$s window before their prompt candidate, (c) if they are followed by any subsequent ID trigger within a 800 $\mu$s window from their prompt candidates, (d) if they are followed by any subsequent ID \& OD trigger within a 200 $\mu$s window from their prompt candidates; (viii) the criteria for removing $^{252}$Cf contamination background, (a) a timing veto requirement if they are accompanied by a prompt candidate of $E_{\textrm{p}}$ > 3 MeV within a 30 s window and a distance of 50 cm for the far detector, (b) a spatial veto requirement for rejecting coincidence pairs in the far detector only if the vertices of their prompt candidates are located in a cylindrical volume of 30 cm in radius, centered at $\textit{x}$ = +12.5 cm and $\textit{y}$ = +12.5 cm and $-$170 < $\textit{z}$ < $-$120 cm. 


\begin{table}[tbp]
	
	\begin{center}
		\begin{tabular*}{0.95\textwidth}{@{\extracolsep{\fill}} c  c  c}\hline\hline
			& n-H & n-Gd\\\hline
			Prompt energy cut & 0.7 < $E_{\textrm{p}}$ < 12.0 MeV & 0.7 < $E_{\textrm{p}}$ < 12.0 MeV\\
			Delayed energy cut & 2.223 $\pm$ 2 $\sigma$ & 6 < $E_{\textrm{d}}$ < 12.0 MeV\\
			Time coincidence ($\Delta t$) & 2 < $\Delta t$ < 400 $\mu$s  & 2 < $\Delta t$ < 100 $\mu$s\\
			Spatial coincidence ($\Delta R$) & < 450 mm & < 2500 mm\\
			$Q_{\textrm{max}}^{\textrm{prompt}}$/$Q_{\textrm{tot}}^{\textrm{prompt}}$ & < 0.07 & < 0.08\\
			$Q_{\textrm{max}}^{\textrm{delayed}}$/$Q_{\textrm{tot}}^{\textrm{delayed}}$ & < 0.06 & < 0.08\\
			\hline\hline
		\end{tabular*}
	\end{center}
	\caption{Comparison of IBD selection criteria for the n-H and n-Gd analysis.}
	\label{tab:IBDSel}
\end{table}

Applying the IBD selection criteria yields 567690 (90747) candidates events with 1.2 < $E_{\textrm{p}}$ < 8.0 MeV 
for a live time of 1546.61 (1397.72) days from 11 August 2011 to 23 April 2017 in the near (far) detector. 
The IBD candidates include remaining backgrounds of correlated or
uncorrelated pairs between the prompt and delayed-like events. The near detector
suffers from a higher cosmogenic background rate because of its shallower
overburden than the far detector. The uncorrelated IBD background comes from accidental coincidence between two randomly correlated events. The prompt-like event is mostly due to ambient $\gamma$-rays from detector materials and surrounding rocks. The correlated IBD backgrounds are produced by fast neutrons, cosmogenic $^9$Li/$^8$He
isotopes, and $^{252}$Cf contamination in the detector target.

\begin{figure}[tbp]
	\centering
	\includegraphics[width=0.7\textwidth]{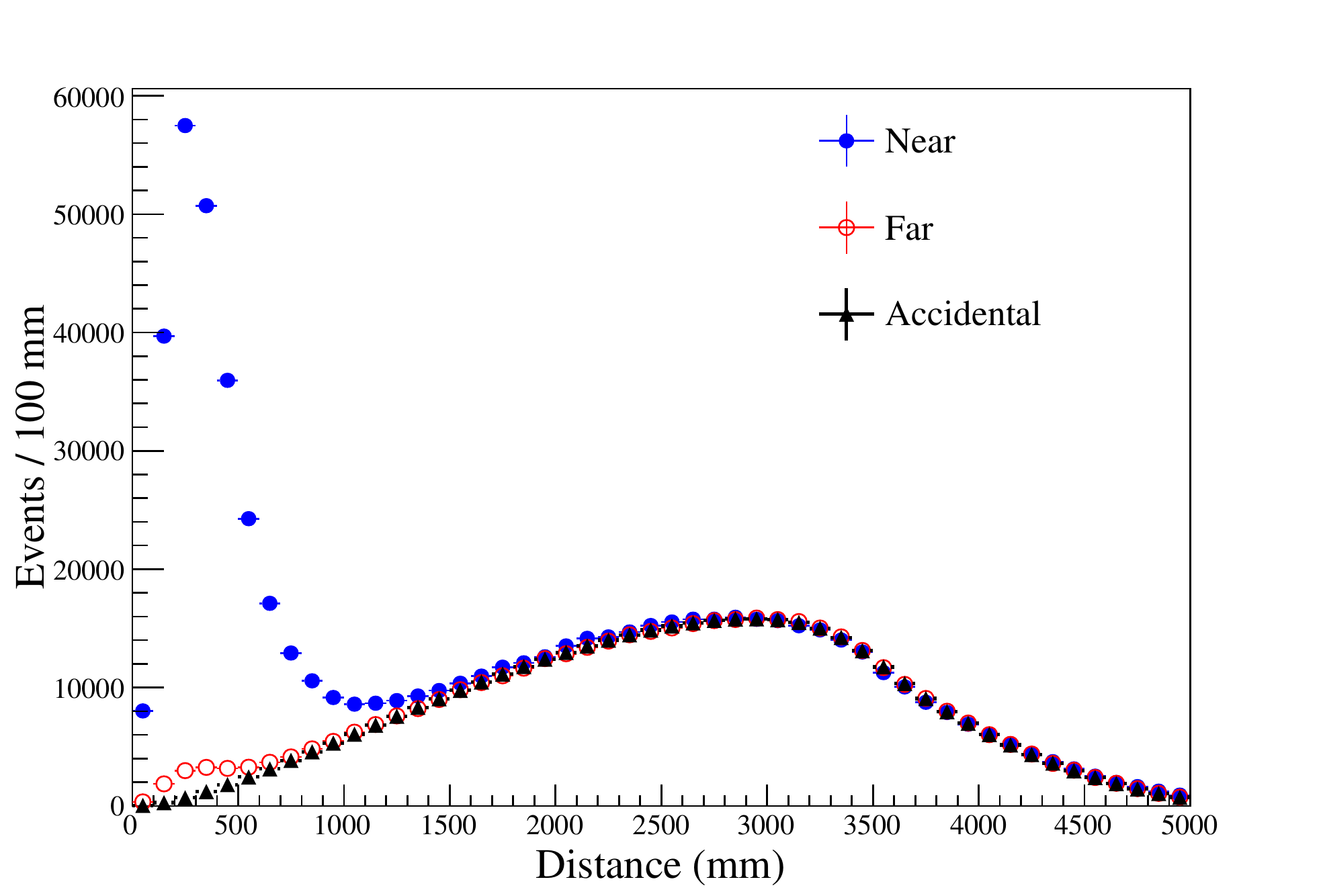}
	\caption{Distance between prompt and delayed candidates. The far data (red)
		is normalized to the near data (blue) at $\sim$~3000 mm. A requirement of $\Delta R <450$ mm is efficient to remove most of the accidental background.
		}
	\label{fig:nHdR}
\end{figure}

\section{Backgrounds}

\subsection{Accidental background}

\begin{figure}[tbp]
	\centering
	\includegraphics[width=0.95\textwidth]{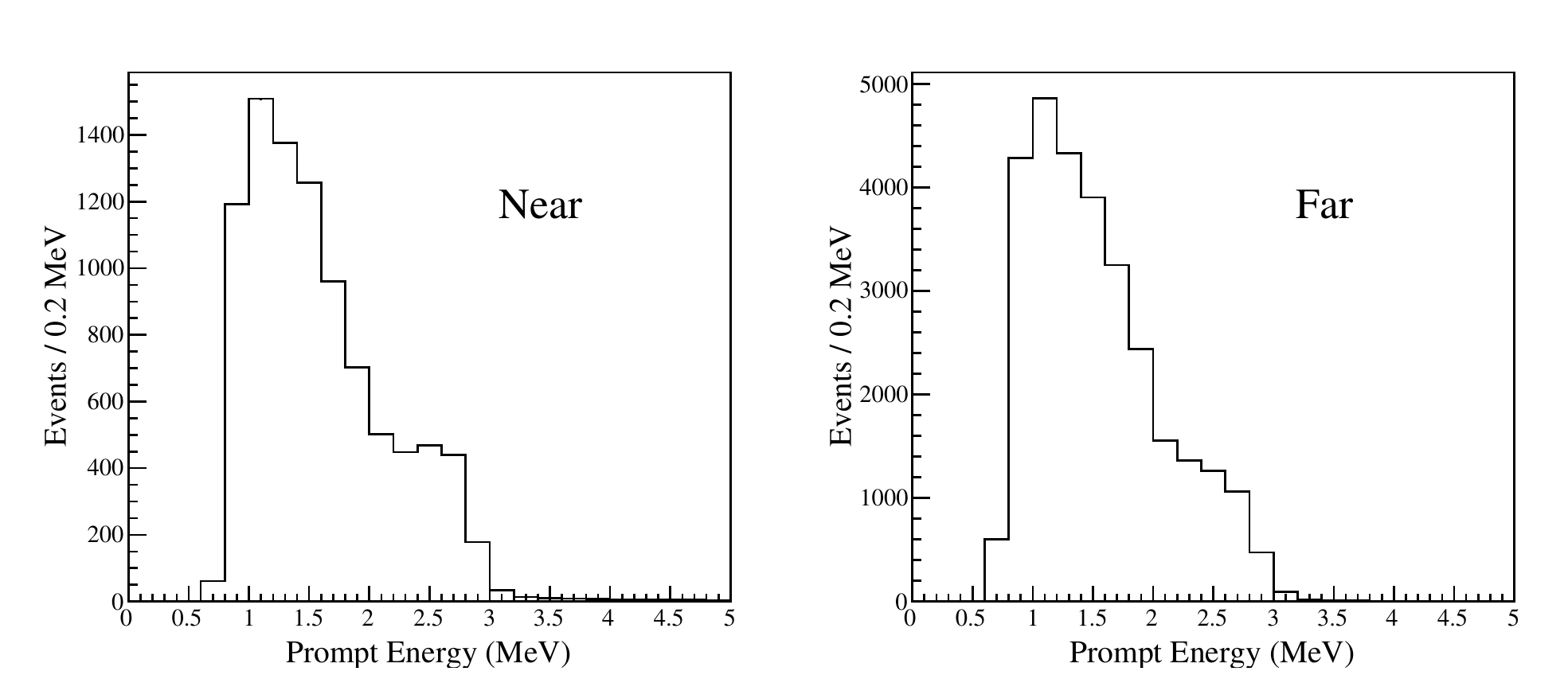}
	\caption{Prompt energy spectra of accidental backgrounds. They are
		normalized to the remaining accidental backgrounds. The background rate of the
		far detector is $\sim$3 times higher than that of the near detector as expected from the higher radioactivity measured in the rock sample of the far site.}
	\label{fig:AcciEn}
\end{figure}

An accidental background comes from the random association of a prompt-like event and a delayed-like neutron capture. The prompt-like events are mainly ambient $\gamma$-rays from the radioactivity in the PMT glasses, LS, and surrounding rock. 
The n-H delayed-like events are overwhelmed by the high-rate
prompt-like events of ambient $\gamma$-rays around 2.2 MeV, unlike the n-Gd
delayed event. A fake delayed event paired with a prompt-like event introduces
significant increase of an accidental background rate in the n-H analysis.
The remaining accidental background rate in the final sample is estimated
by a fit to the $\Delta R$ distribution. Figure \ref{fig:nHdR} shows the $\Delta R$ distributions of IBD signal and accidental background. The rate of random spatial associations in the IBD signal region of $\Delta R$ < 450 mm is estimated by extrapolating from the background dominant region of $\Delta R$ > 2000 mm using $\Delta R$ distribution of accidental background. 
 An accidental background enriched sample is obtained from a requirement of temporal association larger than 1 ms. The prompt energy spectrum of accidental background is obtained from the control sample as shown in Fig. \ref{fig:AcciEn} and consistent with the expectation for the ambient $\gamma$-rays emitted by natural radio-isotopes of $^{40}$K, $^{232}$Th, and $^{238}$U. The estimated accidental background rate is 8.48$\pm$0.01 (21.76$\pm$0.01) events per day in the near (far) detector. The accidental background rate is influenced by external $\gamma$-rays
from the rock surrounding the detector. The accidental background rate of the far
detector is about three times higher than that of the near detector.

\subsection{Fast neutron background}
\begin{figure}[tbp]
	\centering
	\includegraphics[width=0.95\textwidth]{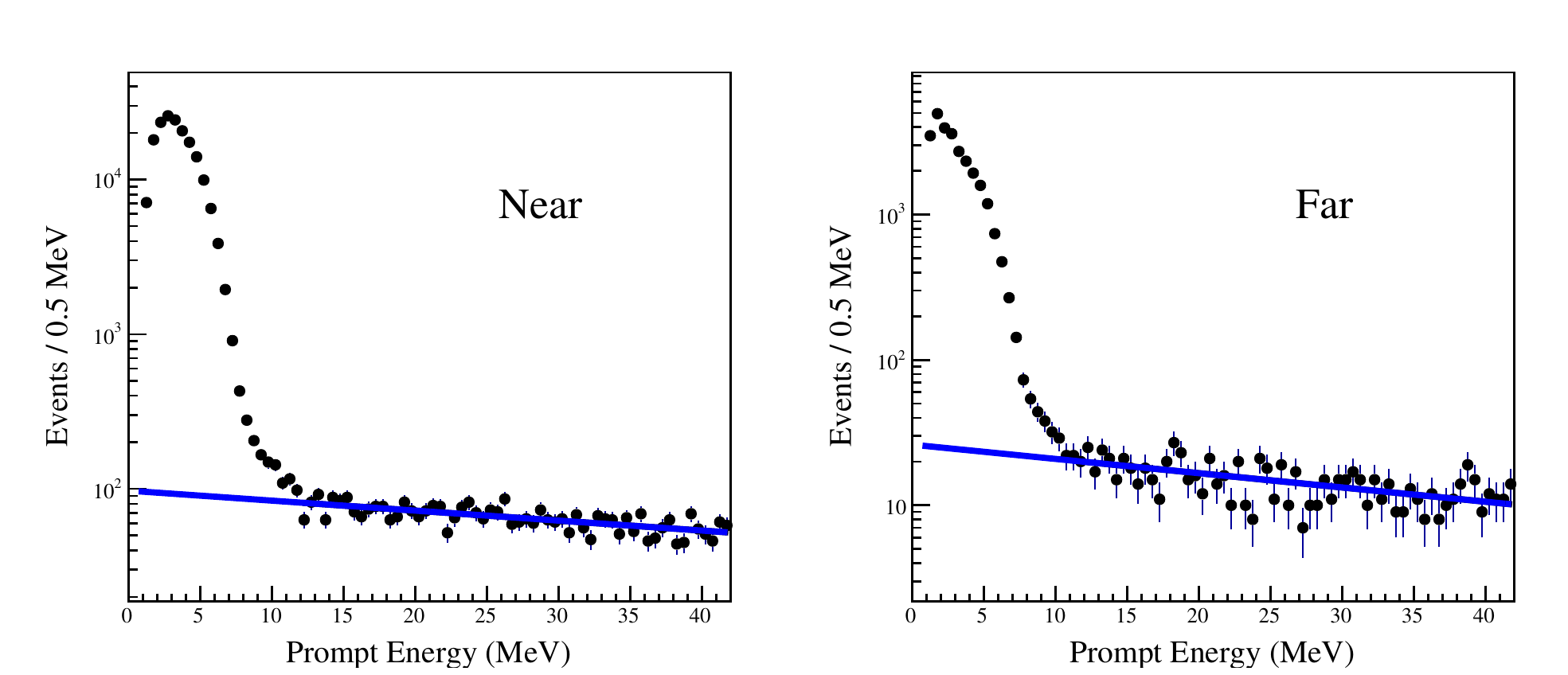}
	\caption{Prompt energy spectra of IBD candidates and fast neutron background. The remaining rate of fast neutron in the IBD candidates is estimated by extrapolating from the background dominant region assuming an exponential spectrum of the background.}
	\label{fig:FnEn}
\end{figure}

The energetic neutrons are produced via spallation when cosmic muons traverse the surrounding rock or the detector. The neutron entering the detector interacts with a proton in LS and produces a recoil proton that generates scintillation lights mimicking a prompt-like event. After loosing kinetic energy through the multiple interactions, the neutron becomes thermalized and captured by H or Gd. The remaining fast neutron background rate is obtained from an IBD candidate sample with the prompt energy extended up to 60 MeV. The observed energy spectrum of the fast neutron background as shown in Fig. \ref{fig:FnEn} exponentially decreases as the prompt energy increases. The fast neutron enriched sample is obtained by selecting IBD candidates that are accompanied by a prompt event of $E_{\textrm{p}}$ > 0.7 MeV within 400 $\mu$s. The energy spectral shape of the fast neutron background in the IBD signal region is confirmed to be exponential from the fast neutron enriched
sample.

The amount of fast neutron background remaining in the IBD candidates is estimated by extrapolating from the background dominant energy region of $E_{\textrm{p}}$ > 12 MeV, assuming an exponential slope as shown in Fig. \ref{fig:FnEn}.
The fast neutron background rate is 3.16$\pm$0.12 (0.80$\pm$0.12) events per day in the near (far) detector.

\subsection{Cosmogenic $^9$Li/$^8$He background}

\begin{figure}[tbp]
	\centering
	\includegraphics[width=0.7\textwidth]{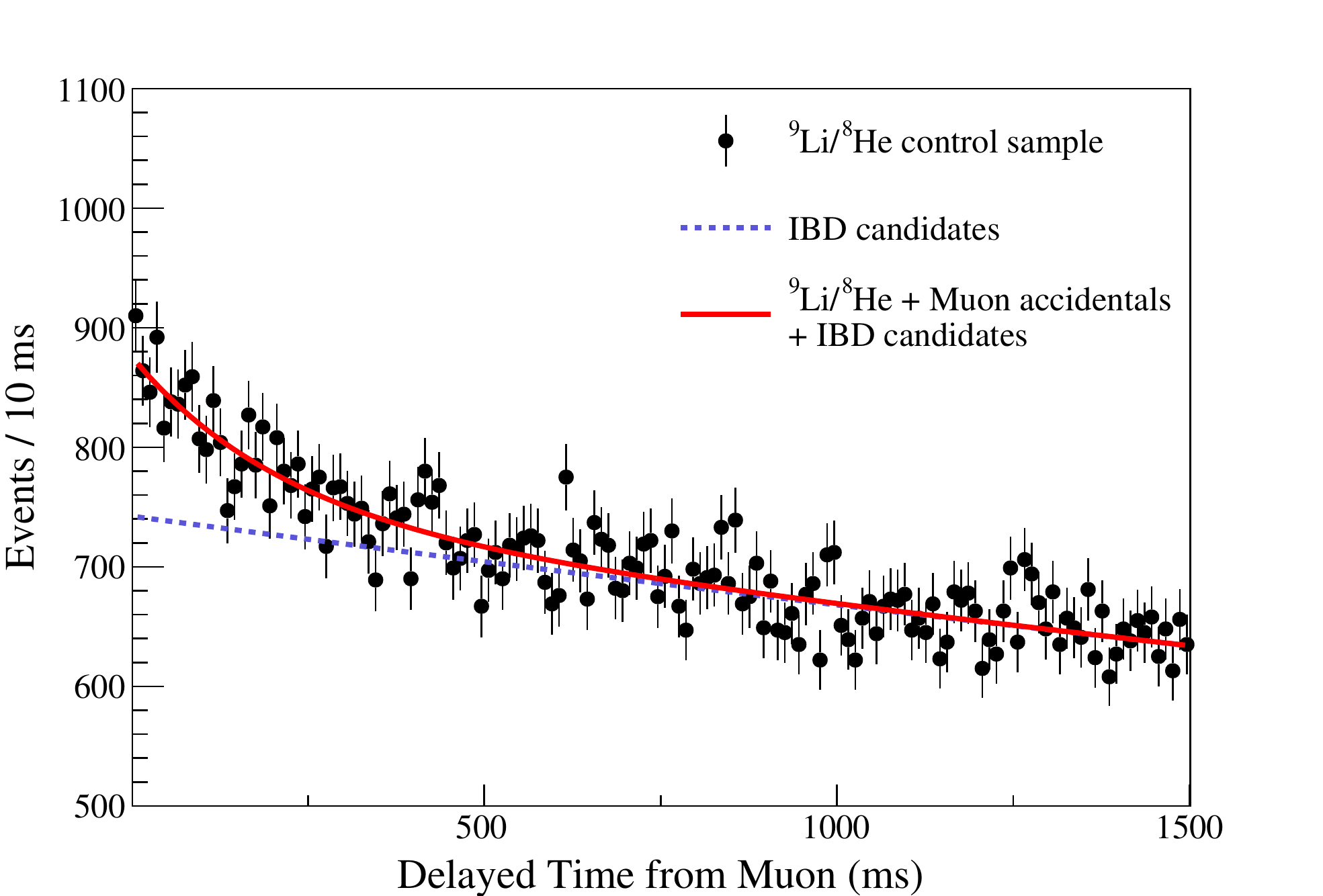}
	\caption{Delayed time distribution of IBD-like pairs from their preceding
		energetic muons in the near detector. There are three components of
		muon-induced accidental background, $^9$Li/$^8$He background, and IBD signal. The $^9$Li/$^8$He background is clearly seen within 500 ms after an energetic muon.}
	\label{fig:MuTimeDiff}
\end{figure}

\begin{figure}[tbp]
	\centering
	\includegraphics[width=0.95\textwidth]{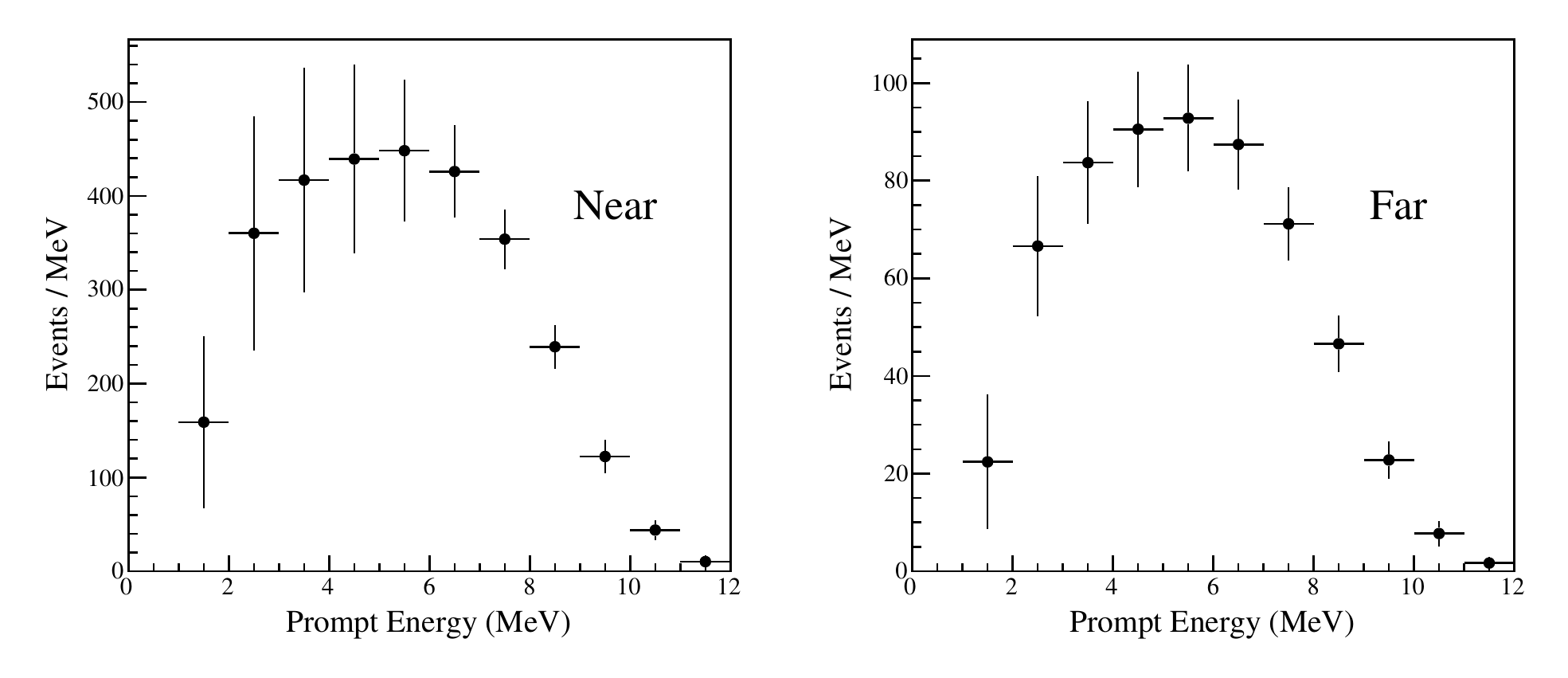}
	\caption{Measured prompt energy spectra of $^9$Li/$^8$He background using their enriched samples of 2000 live days, after subtracting the IBD signal and the muon-induced accidental background. The error bars represent the statistical error of the enriched sample and the fit uncertainty of delayed time distribution.
	}
	\label{fig:LiHe_en}
\end{figure}

\begin{figure}[tbp]
	\centering
	\includegraphics[width=0.7\textwidth]{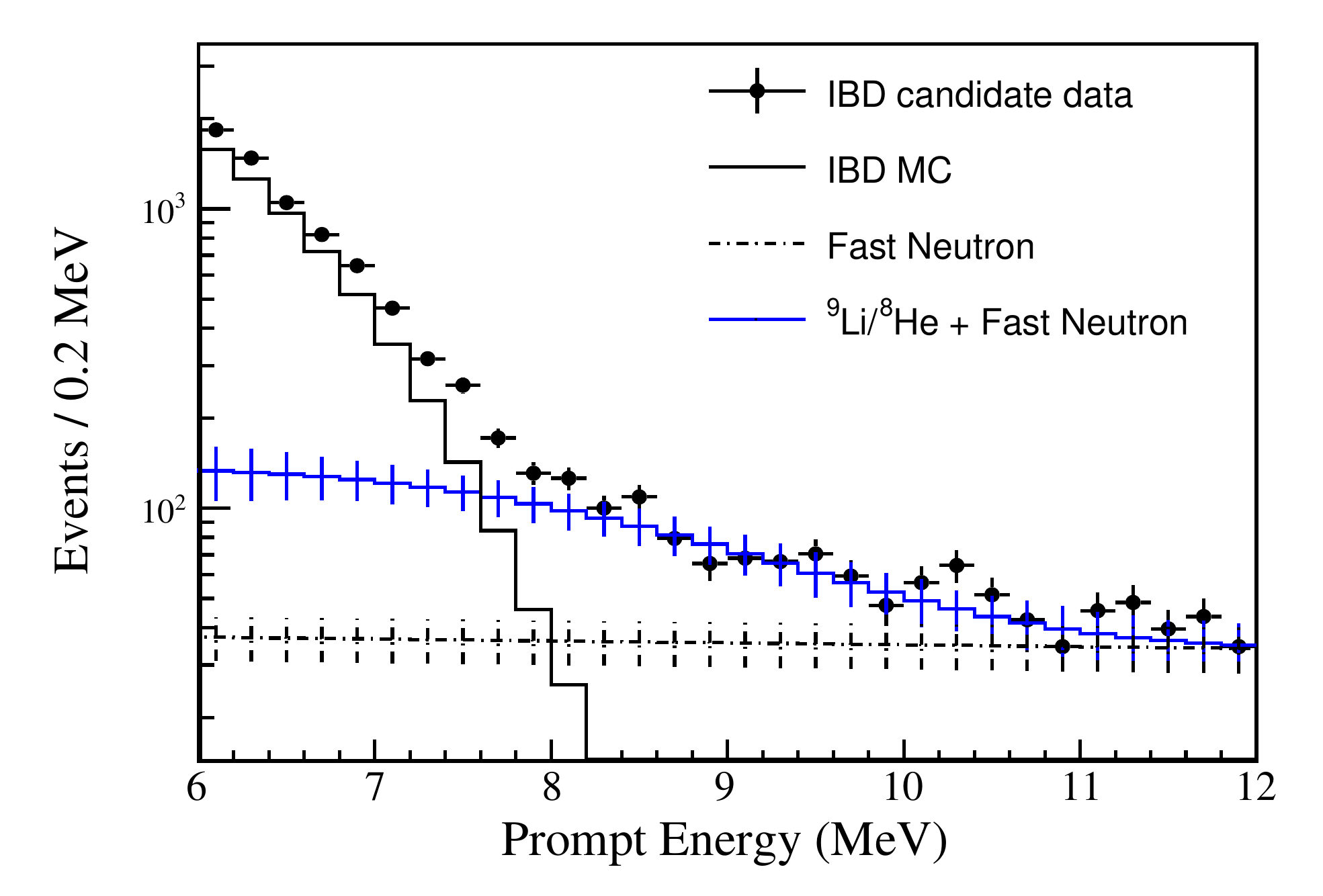}
	\caption{Estimation of remaining $^9$Li/$^8$He background rate in the IBD candidate sample of the near detector. The background rate in the signal region is obtained by extrapolating from the background rate that is measured in the background dominant region of $E_{\textrm{p}}$ > 8 MeV using the measured background spectrum.}
	\label{fig:LiHefit}
\end{figure}

The unstable isotopes of $^9$Li/$^8$He are produced by interaction of cosmic muon with carbon in LS \cite{Muon1,Muon2,Muon3}. Their production cross section increases with the muon energy. The isotopes subsequently decay with emitting an electron and a neutron and mimic the IBD signal. The $^9$Li/$^8$He background event occurs with a measured mean decay-time of $\sim$250 ms after a cosmic muon passes through the detector. The distribution of delayed time between an energetic muon and a subsequent IBD-like pair is shown in Fig. \ref{fig:MuTimeDiff}. The delayed time distribution consists of three components based on their observed spectra. The shortest decay time component is a muon-induced accidental background up to 10 ms after a preceding muon. The accidental IBD-like pair comes from neutrons produced by an energetic muon and randomly associated prompt-like events. The medium decay time component following the
shortest one is the $^9$Li/$^8$He background with their lifetimes of 267.2 and 171.2 ms,
respectively. The longest decay time component is the IBD signals temporally
uncorrelated with muon events. The $^9$Li/$^8$He background is enriched within 400 ms (500 ms) from muon of $E_{\mu}$ > 1.6 GeV ($E_{\mu}$ > 1.5 GeV) for the near (far) detector. As shown in Fig. \ref{fig:LiHe_en}, the energy spectrum of $^9$Li/$^8$He background is obtained by subtracting IBD candidates and muon-induced accidental background from the $^9$Li/$^8$He enriched sample. The background rate in the signal region of $E_{\textrm{p}}$ < 8 MeV is estimated by extrapolating from the background dominant region of $E_{\textrm{p}}$ > 8 MeV using the measured $^9$Li/$^8$He background spectrum, the measured fast neutron background, and the MC IBD expectation as shown in Fig. \ref{fig:LiHefit}. The $^9$Li/$^8$He background rate is 6.49$\pm$0.49 (1.71$\pm$0.21) events per day in the near (far) detector.

\subsection{$^{252}$Cf background}
A small amount of $^{252}$Cf neutron source was accidentally introduced into the target of both detectors during detector calibration in October, 2012. The O-ring in the acrylic container surrounding the radioactive source became loose due to its aging to cause loose seal and tiny leak of $^{252}$Cf into the detector targets.
%
A stringent multiplicity requirement of no trigger or no event near an IBD candidate eliminates most of multiple neutron events coming from the $^{252}$Cf contamination. The requirement is applied differently to the near and far detectors because of much less $^{252}$Cf contamination in the near detector. After applying the requirement, the $^{252}$Cf contamination background becomes negligible for this n-H analysis because almost all of neutrons coming from contamination only in the target region are captured by Gd. The remaining $^{252}$Cf background rate is 0.095$\pm$0.018 events per day only in the far detector and no remaining $^{252}$Cf contamination background events are observed in the near detector.
 
The total background rate is estimated to be 18.13$\pm$0.51 (24.37$\pm$0.24) events per day in the near (far) detector, respectively.
The total background fractions are 4.94$\pm$0.14\% in the near detector, and 37.53$\pm$0.38\% in the far detector. 
The observed rates of IBD and background are summarized in Table ~\ref{tab:bkg}.

\begin{table}[tbp]
	
	\begin{center}
		\begin{tabular*}{0.7\textwidth}{@{\extracolsep{\fill}} l  r  r}\hline\hline
			Detector & Near & Far\\\hline
			IBD rate & 367.05$\pm$0.49 & 64.92$\pm$0.22\\
			After background subtraction & 348.92$\pm$0.70 & 40.55$\pm$0.33\\
			Total background rate & 18.13$\pm$0.51 & 24.37$\pm$0.24\\ 
			DAQ live time (days) & 1546.61 & 1397.72\\
			\hline
			Accidental rate & 8.48$\pm$0.01 & 21.76$\pm$0.01\\
			Fast neutron rate & 3.16$\pm$0.12 & 0.80$\pm$0.12\\
			$^9$Li/$^8$He rate & 6.49$\pm$0.49 & 1.71$\pm$0.21\\
			$^{252}$Cf contamination rate &  & 0.095$\pm$0.018
			\\\hline\hline
		\end{tabular*}
	\end{center}
	\caption{Observed rates of IBD candidates and remaining backgrounds per day for 1.2 < $E_{\textrm{p}}$ < 8.0 MeV.}
	\label{tab:bkg}
\end{table}

\section{Systematic uncertainties}
\subsection{Detection efficiency uncertainty}

The detection efficiency is almost the same for the near and far detectors because of their identical construction and performances. However, there might be slight differences in detection even between two identical detectors. The detection efficiencies of several selection criteria that are applied to both near and far detectors are investigated. There are two types of systematic uncertainties, namely correlated and uncorrelated between both detectors. The correlated uncertainty is common to both detectors and thus cancelled out for a far-to-near ratio measurement. By contrast, the uncorrelated uncertainty is not cancelled out for both detectors.
The total uncorrelated uncertainty of detection efficiency is included in the measurement of $\theta_{13}$ value. Therefore, identical performances of the near and far detectors minimize the uncorrelated uncertainties and allows cancellation of the correlated systematic uncertainties for the ratio measurement. A control data sample is basically used to study the detection efficiencies of the near and far detectors. When a control data sample is not available, MC is used instead. An uncorrelated relative uncertainty of detection efficiency is estimated by comparing the difference between both detectors. In this section detection efficiencies and their systematic uncertainties for the IBD selection are described and presented.

The detection efficiency of H capture fraction is calculated based on the ratio of neutron captures on H relative to total neutron captures of the IBD signal.
In the target region, neutron can be captured by H, C, or Gd. The fraction of neutron capture by C is less than 0.1\% and can be neglected. The H capture fraction and uncorrelated uncertainty in the target plus GC region are determined to be 69.42\% and 0.04\%, respectively. The efficiencies for various selection criteria and the total detection efficiency are calculated based on the events only in which neutrons are
captured by hydrogen.

A main trigger for an IBD candidate event requires ID $N_{\textrm{hit}}$ > 90 within a 50 ns time window. For MC, a requirement of $N_{\textrm{hit}}$ > 84 is imposed to make data and MC equivalent \cite{RENOPRD}. The trigger efficiency is estimated from the IBD signal loss due to the requirement and to be 78.79$\pm$0.01\% using the MC. The IBD signal loss due to the trigger requirement takes place when an event occurring in the GC’s outer layer emits minimal scintillating lights, leaves most energy deposit in the buffer region of non-scintillating oil, and generates insufficient PMT hits for a trigger. Uncorrelated systematic uncertainty is estimated as 0.02\% from the difference between near and far efficiencies.

The efficiency of the prompt energy requirement is obtained by
calculating the fraction of events in the region of 1.2 < $E_{\textrm{p}}$ < 8 MeV relative to total IBD events using MC and estimated as 97.83$\pm$0.01\%. The uncorrelated systematic uncertainty is obtained by varying the energy threshold according to the energy scale difference of 0.5\% between the near and far detectors. The relative energy scale difference between the detectors is estimated by comparing the charge-to-energy conversion functions from various radioactive calibration sources and is found to be less than 0.5\%.
The uncorrelated uncertainty is estimated to be 0.08\%.

The efficiency of delayed energy cut is determined by the fraction of delayed events in the region (2.223$\pm$2$\sigma _d$) MeV out of total delayed events of neutron capture on H where $\sigma _d$ is the delayed energy resolution. 
A control sample for the efficiency measurement of delayed events is obtained by requiring conditions of E$_p$ > 4 MeV and $\Delta$R < 300 mm to remove 
accidental backgrounds.
There are almost no events above 3 MeV in the delayed energy distribution of Fig. \ref{fig:nHs2}. For the energy range below 3 MeV, the efficiency is estimated to be 86.71$\pm$0.16\%. 
The uncorrelated systematic uncertainty is estimated to be 0.16\% by changing the delayed energy requirement by $\pm$0.5\%, the energy scale difference between the near and far detectors.

\begin{figure}[tbp]
	\centering
	\includegraphics[width=0.95\textwidth]{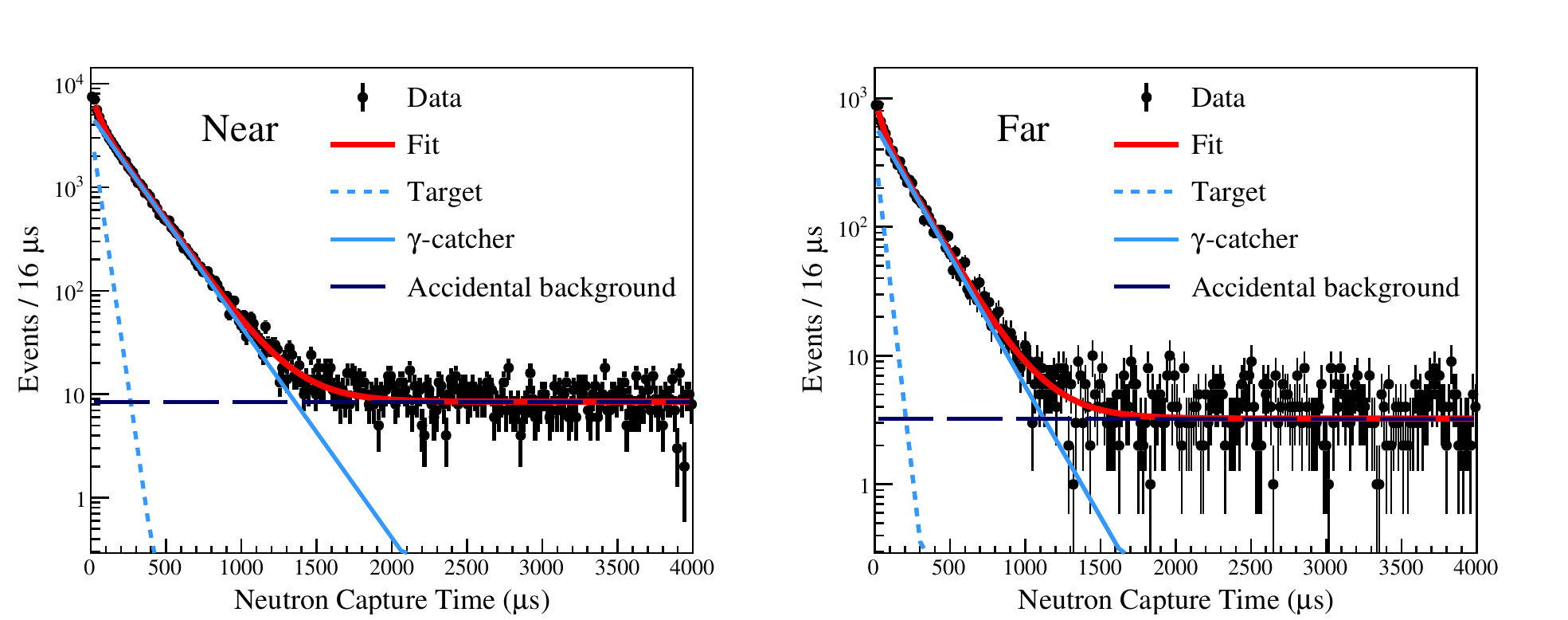}
	\caption{
		Time distribution of neutron capture on hydrogen. There are three components of the IBD signals in the target and $\gamma$-catcher regions and the accidental background. The red solid line is the best fit to the data. The blue dotted (solid) line represents the fitted capture time of the IBD signal in the target ($\gamma$-catcher) region. The black dashed line corresponds to the capture time of the accidental background.}
	\label{fig:Dt_eff}
\end{figure}
 
The efficiency of the time coincidence requirement is obtained by the fraction of IBD events with 2 < $\Delta t$ < 400 $\mu$s out of total IBD events. In order to obtain this efficiency, E$_p$ > 4.5 MeV requirement is applied in order to reduce accidental backgrounds. The neutron capture time distribution of the IBD signal sample is shown in Fig. \ref{fig:Dt_eff}. The data is fitted by a model of two exponential functions and a constant term. The mean neutron capture time is $\sim$200 $\mu$s for hydrogen and, in contrast, $\sim$30 $\mu$s for 0.1\% Gd-loaded LS due to high capture cross-section of Gd \cite{Gdcross}. Two capture time components are found for the n-H IBD signals in the target and $\gamma$-catcher regions. The third component is for the accidental background. The efficiency is estimated to be 85.30$\pm$0.12\% from two exponential capture time distributions of the IBD signals. The Gd concentration difference between the near and far detectors is less than 0.1\% as every batch of Gd-loaded LS is equally divided into both detectors during detector construction. The Gd concentration difference between two detectors results in the efficiency difference of the time coincidence requirement, obtained from the MC. The uncorrelated uncertainty of this requirement is estimated to be 0.04\%.

\begin{figure}[tbp]
 	\centering
 	\includegraphics[width=0.95\textwidth]{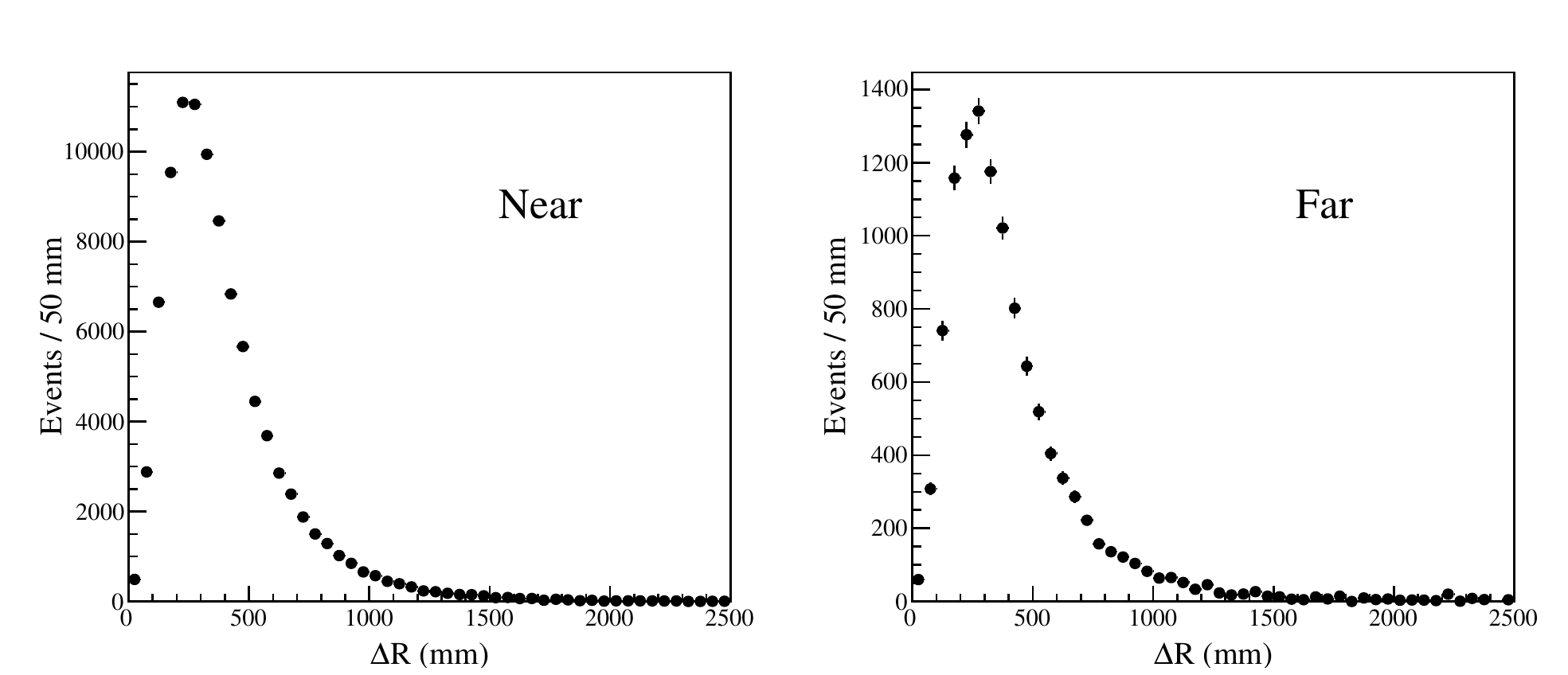}
 	\caption{$\Delta R$ distribution of IBD candidates with almost no accidental background.}
 	\label{fig:DR_eff}
 \end{figure}

The efficiency of the spatial correlation requirement is measured by the fraction of IBD candidates with $\Delta R$ < 450 mm out of total IBD events using a control sample selected by a prompt energy requirement of E$_p$ > 4 MeV in order to minimize the accidental background. Figure \ref{fig:DR_eff} shows the $\Delta R$ distribution of IBD candidates with almost no accidental background. The efficiency of the $\Delta R$ requirement is estimated to be 70.49$\pm$0.13\%. The uncorrelated uncertainty is estimated to be 0.09\% from the efficiency difference between the near and far detectors.
 
 \begin{table}[tbp]		
 	\centering
 	\begin{tabular*}{0.7\textwidth}{@{\extracolsep{\fill}} l c c}
 		\hline\hline
 		&Efficiency & Uncorrelated \\\hline
 		H capture fraction & 69.42\%&0.04\% \\
 		Trigger efficiency & 78.79\%& 0.01\% \\
 		Prompt energy cut &97.83\%& 0.08\% \\
 		Delayed energy cut &86.71\%& 0.16\% \\
 		Time coincidence cut &85.30\%& 0.04\% \\
 		Spatial correlation cut &70.49\%& 0.09\% \\			
 		$Q_{\textrm{max}}^{\textrm{prompt}}$/$Q_{\textrm{tot}}^{\textrm{prompt}}$&92.22\%& 0.08\% \\
 		$Q_{\textrm{max}}^{\textrm{delayed}}$/$Q_{\textrm{tot}}^{\textrm{delayed}}$&87.45\%& 0.08\% \\
 		\hline
 		Total detection efficiency &32.46\%& 0.24\% \\
 		\hline\hline
 	\end{tabular*}
 	\caption{Detection efficiencies and their uncorrelated uncertainties of selection criteria for IBD signal. The efficiency is calculated as the statistical error weighted mean of the near and far measured values. }
 	\label{tab:DetEff}
 \end{table}
 
Events with $Q_{\textrm{max}}^{\textrm{prompt}}$/$Q_{\textrm{tot}}^{\textrm{prompt}}$ > 0.08 are rejected.
The efficiency of the $Q_{\textrm{max}}^{\textrm{prompt}}$/$Q_{\textrm{tot}}^{\textrm{prompt}}$ requirement is obtained using IBD candidates with the stringent conditions of E$_p$ > 4 MeV and $\Delta R$ < 350 mm in order to remove the background events.   
Events with $Q_{\textrm{max}}^{\textrm{prompt}}$/$Q_{\textrm{tot}}^{\textrm{prompt}}$ > 0.08 is determined by extrapolating from the region of $Q_{\textrm{max}}^{\textrm{prompt}}$/$Q_{\textrm{tot}}^{\textrm{prompt}}$ < 0.08.  
In addition, an expected shape of its distribution is also confirmed by MC.   
The efficiency of $Q_{\textrm{max}}^{\textrm{prompt}}$/$Q_{\textrm{tot}}^{\textrm{prompt}}$ requirement is estimated as 92.22$\pm$0.29\%. 
The uncorrelated uncertainty is estimated to be 0.08\% from the efficiency difference between the near and far detectors.

The requirement of $Q_{\textrm{max}}^{\textrm{delayed}}$/$Q_{\textrm{tot}}^{\textrm{dalayed}}$ < 0.06 is applied for selecting IBD candidates. The method of obtaining this efficiency is identical to that of prompt $Q_{\textrm{max}}$/$Q_{\textrm{tot}}$ requirement. The requirement efficiency is estimated as 87.45$\pm$0.24\%. Its uncorrelated uncertainty is estimated to be 0.08\% from the efficiency difference between the near and far detectors.

The detection efficiencies of several selection criteria applied to both near and far detectors are listed in Table \ref{tab:DetEff}. 
The total detection efficiency is estimated to be 32.46\%, and the total uncorrelated uncertainty is 0.24\% using data or MC. 
The fraction of detection efficiency ($\epsilon$) to its uncorrelated uncertainty ($\Delta \epsilon$) was calculated to be 0.73\%, and used in the measurement of $\theta_{13}$ value.
The main contributions to the uncorrelated uncertainty come from different efficiencies between the two detectors associated with the delayed energy requirement. 

Among the IBD selection criteria, the muon and multiplicity timing veto requirements are applied differently to the near and far detectors because of different overburden and surrounding environments at near and far sites. Therefore, the signal loss due to a timing veto requirement differs between two detectors depending on their muon or trigger rates. The fraction of IBD signal loss by the muon timing veto is determined to be 21.56\% (11.40\%) for the near (far) detector. The IBD signal loss due to multiplicity timing veto requirement is 25.29$\pm$0.04\% (11.14$\pm$0.03\%) for the near (far) detector. The signal loss of each criteria is summarized in Table \ref{tab:sigloss}.

\begin{table}[tbp]
	
	\begin{center}
		\begin{tabular*}{1.0\textwidth}{@{\extracolsep{\fill}} l  c  c}\hline\hline
			Timing veto criteria & Near (\%) & Far (\%)\\\hline
			Timing criteria associated with muon & 21.564$\pm$0.002 & 11.398$\pm$0.002\\
			IBD candidate accompanied by any trigger&18.744$\pm$0.001\newline & 5.089$\pm$0.001\\
			within 500 $\mu$s preceding time window & & \\
			IBD candidate accompanied by ID and ID\&OD trigger & 2.013$\pm$0.001 & 1.242$\pm$0.001\\
			within 600 $\mu$s preceding time window & & \\
			IBD candidate accompanied by ID\&OD trigger & 2.303$\pm$0.058 & 0.171$\pm$0.037\\
			within 200 $\mu$s subsequent time window & &\\
			IBD candidate accompanied by ID-only trigger  & 3.921$\pm$0.001 & 5.028$\pm$0.001\\
			within 800 $\mu$s subsequent time window & &\\
			IBD candidate accompanied by prompt candidate & & 0.997$\pm$0.029\\
			(> 3 MeV) within 30 s subsequent time window and 50 cm & &\\
			\hline
			Combined IBD signal loss& 41.406$\pm$0.035 &22.135$\pm$0.045\\
			\hline\hline
		\end{tabular*}
	\end{center}
	\caption{IBD signal loss due to timing veto criteria. The loss is determined
		by trigger or muon rates.}
	\label{tab:sigloss}
\end{table}

\subsection{Background uncertainty}

The background uncertainty is an essential part in determining the error of $\theta_{13}$. The background estimation has already been described in the previous section. In the rate-only analysis, the uncertainties of the remaining
background rate and spectral shape, as listed in Table \ref{tab:bkg}, are used in the measurement of $\theta_{13}$. Among all backgrounds of the n-H analysis, the largest rate comes from the accidental background and the largest uncertainty from the $^9$Li/$^8$He background.

\subsection{Reactor related uncertainty}

The antineutrino flux is crucial in determining the $\theta_{13}$ value and
suffers from the reactor related uncertainties. The expected rate of reactor
$\overline{\nu}_e$ during physics data taking depends on the thermal power output, fission fractions of four isotopes, energy released per fission, and IBD capture cross-section. The sources of uncorrelated uncertainties of the near and far detectors related to reactors are baseline distance, reactor thermal power, and fission fraction. The positions of the two detectors and six reactors are surveyed by the global positioning system and total stations. The baseline distance between the detectors and reactors can be measured with an accuracy of less than 10 cm. 
The uncertainty of baseline distance is 0.03\%, and can be neglected for
determining the $\theta_{13}$ value. The uncertainties of thermal power output are 0.5\% per core \cite{ReacUnc}. The relative fission contribution of the four isotopes have 4$\sim$10\% uncertainties during the fuel cycle \cite{IsoUnc}. The uncertainties of fission fraction contribute 0.7\% of the neutrino yield per core to the uncorrelated uncertainty \cite{IsoUnc}. The combined uncorrelated uncertainty of reactor flux is estimated as 0.9\% and used in the measurement of $\theta_{13}$ value.



\section{Results and summary}

\begin{figure}[tbp]
	\centering
	\includegraphics[width=0.95\textwidth]{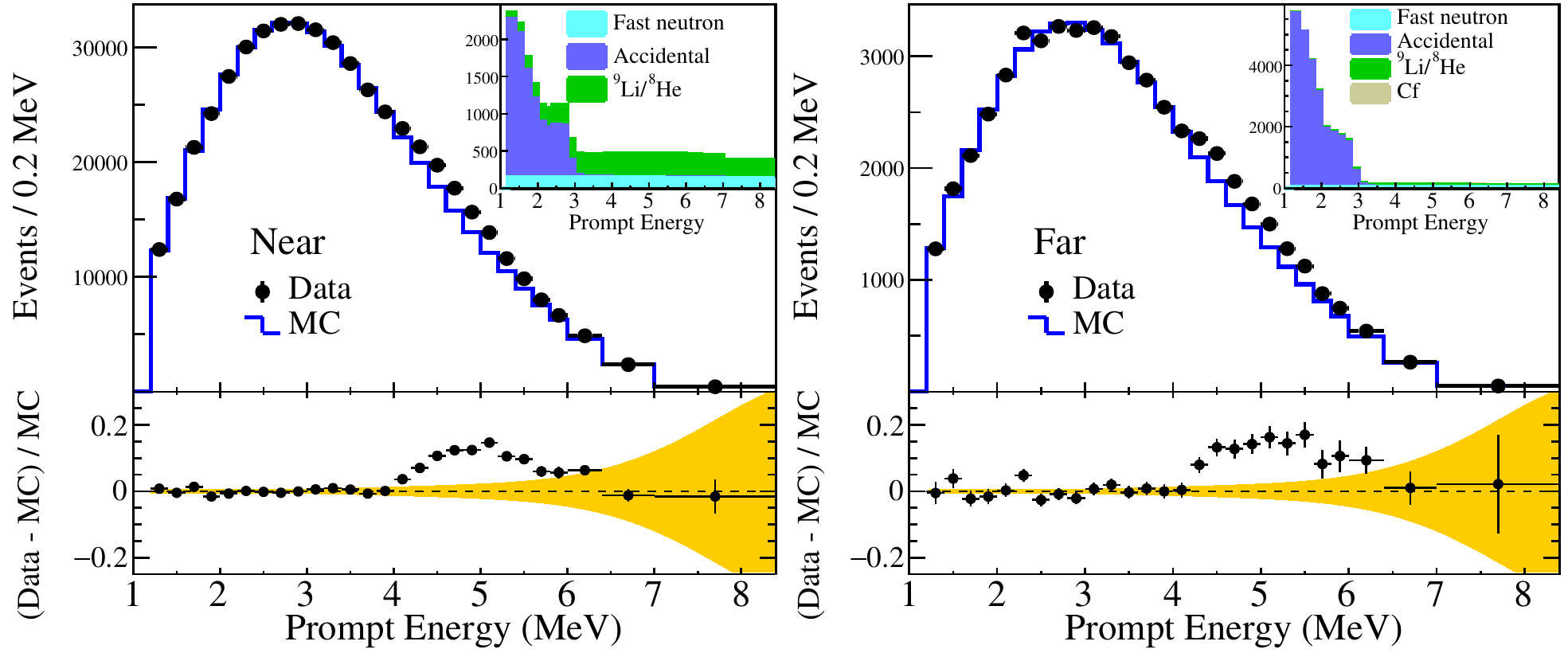}	
	\caption{Observed and expected prompt-energy spectra in the near and far detectors. The Huber and Mueller model is used for the expected spectra. The remaining backgrounds are shown in the insets. The fractional difference between data and MC is shown in the lower panel. The yellow shaded bands represent the uncertainties of expected spectra. A spectral deviation from the expectation around 5 MeV is larger than the uncertainty of an expected spectrum from the reactor antineutrino model \cite{Exp4,Exp5}.}
	\label{fig:5mev}
\end{figure}

The energy spectra of the observed IBD prompt events after background subtraction are shown in Fig. 17 \ref{fig:5mev}.
The MC expected energy spectra are obtained from a reactor neutrino model \cite{Exp4,Exp5} and the best-fit oscillation results. The n-H data also show clear discrepancy between the observed and MC predicted spectra around 5 MeV in both detectors. The RENO experiment first reported the unexpected 5 MeV excess in 2014 using 800 live days of n-Gd data \cite{RENO5MeV}. The excess is found to be consistent with coming from reactors and amounts to ~3\% of the total observed IBD events in both detectors. It would be interesting to find if the excess has a correlation with a particular isotope of reactor fuel composition \cite{RENOPRLFuel}. 

 The oscillation amplitude of neutrino survival probability is extracted from the observed reactor $\overline{\nu}_e$ rates. Even with the unexpected shape in the observed reactor neutrino spectrum, the oscillation amplitude can be determined from a fit to the measured far-to-near ratio of IBD rate. The 5 MeV excess does not affect the determination due to its cancellation in the ratio measurement using the identical near and far detectors.

For determining the mixing angle $\theta_{13}$, a $\chi^2$ with pull parameters associated with uncorrelated systematic uncertainties is minimized by varying the oscillation amplitude and pull parameters \cite{Chi}. The following $\chi^2$ function as used in the n-Gd analysis \cite{RENOPRL1} is applied for the determination:

\begin{eqnarray}
\chi ^2 = \frac{(O^{F/N}-T^{F/N})^2}{U^{F/N}} + \sum_{d=N,F}( \frac{b_{d}}{\sigma _{bkg}^d} )^2
+ \sum_{r=1\sim 6}( \frac{f_{r}}{\sigma _{flux}} )^2
+ ( \frac{\varepsilon}{\sigma _{eff}} )^2, 
\end{eqnarray}

where $O^{F/N}$ is the far-to-near ratio of observed IBD candidates, $U^{F/N}$ is the statistical uncertainty of $O^{F/N}$, and $T^{F/N}$ is the far-to-near ratio of expected IBD events including reactor neutrino flux, IBD cross-section, survival probability and detection efficiency. Index $d$ stands for the near ($N$) and far ($F$) detectors. 
The systematic uncertainty sources are embedded by pull parameters ($b^d, f_r$ and $\varepsilon$) with associated uncertainties
 ($\sigma _{bkg}^d, \sigma _{flux}$ and $\sigma _{eff}$).
The pull parameters allow variation from the expected far-to-near ratio of IBD events within their corresponding systematic uncertainties. 
The uncorrelated reactor uncertainty ($\sigma_{flux}$) is 0.9\%, the uncorrelated detection uncertainty ($\sigma_{eff}$) is 0.73\%, 
and the background uncertainty ($\sigma _{bkg}^d$) is presented in Table ~\ref{tab:bkg}. 

The observed reactor $\overline{\nu}_e$ rate only is used to extract the oscillation amplitude of neutrino survival probability. We observed a clear deficit in the observed rate, 6.8\% for the far detector and 1.1\% for the near detector with respect to the expected one, 
indicating a definitive observation of reactor antineutrino disappearance consistent with neutrino oscillation. 
Using the deficit information, the obtained best-fit value is 

\begin{figure}[tbp]
	\centering
	\includegraphics[width=0.7\textwidth]{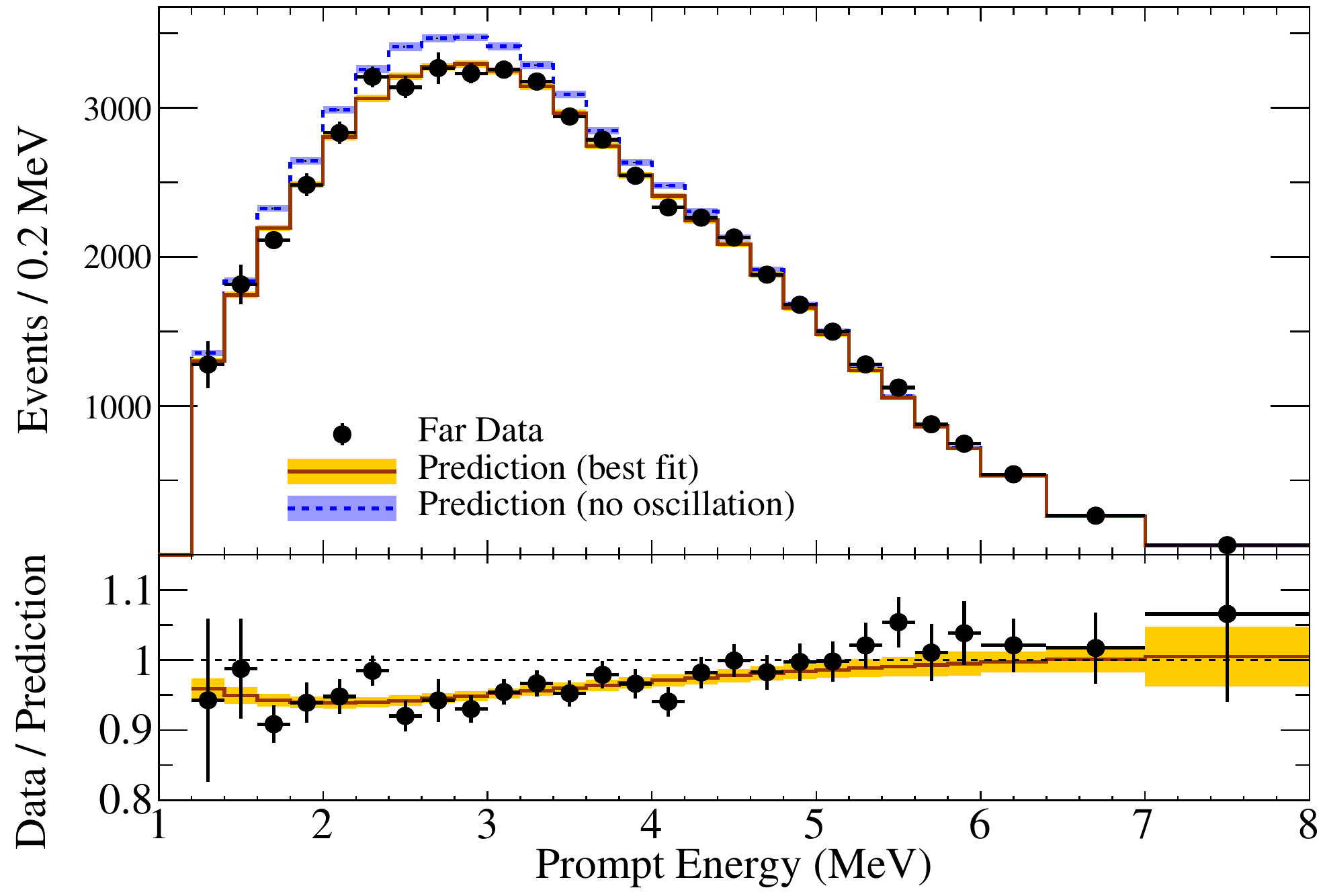}
	
	\caption{Top: Comparison of the observed (dots) and no-oscillation predicted (blue shaded histogram) IBD prompt spectra in the far detector. The no-oscillation prediction is obtained from the measurement in the near detector. The prediction from the best-fit oscillation amplitude is also shown (yellow shaded histogram). Bottom: Ratio of observed spectrum in the far detector to the no-oscillation prediction (dots), and the ratio from the MC simulation with the best-fit results folded in (shaded band). Errors include the statistical and background subtraction uncertainties.}
	\label{fig:eratio}
\end{figure}

\begin{center}
sin$^{2}$2$\theta _{13}$ = 0.087 $\pm $0.008 (stat.) $\pm$ 0.014 (syst.),
\end{center}

where the world average value of |$\Delta m_{ee}^2$| = $(2.502 \times 10^{-3}$ eV$^2)$ is used \cite{PDG2017}. 
This value is consistent with the previous measurement of n-Gd result, sin$^{2}$2$\theta _{13}$=0.0896$\pm$0.0068 within their uncertainties while the systematic uncertainty is about twice larger than that of the n-Gd result \cite{RENOPRL3}.
The error and its fraction of sin$^{2}$2$\theta _{13}$ by each component can be obtained using the pull terms of the $\chi ^2$ equation and are summarized in Table \ref{Tab:sin syst}. The systematic uncertainties of detection efficiency and backgrounds mostly contribute to the systematic error of 1.5 times larger than the statistical error.
%
Figure \ref{fig:eratio} shows the background-subtracted, observed IBD prompt energy spectrum at the far detector compared to the one expected with no oscillation and the one with the best-fit oscillation parameters at the far detector.   

\begin{table}[H]
	\begin{center}
		\begin{tabular}{ccc}\hline\hline
			Uncertainties & Error & Error Fraction (\%)\\\hline
			Reactor & 0.003 & 4.8\\
			Detection Efficiency& 0.010  & 52.6\\
			Backgrounds &  0.009 & 42.6
			\\\hline\hline
			Combined & 0.014 &
			\\\hline\hline
		\end{tabular}
	\end{center}
	\caption{Systematic errors from uncertainty sources. The dominant source of the systematic error for $\sin^2 2\theta _{13}$ is the uncertainty of the remaining background.
	}
	\label{Tab:sin syst}
\end{table}

\begin{figure}[tbp]
	\centering
	\includegraphics[width=0.6\textwidth]{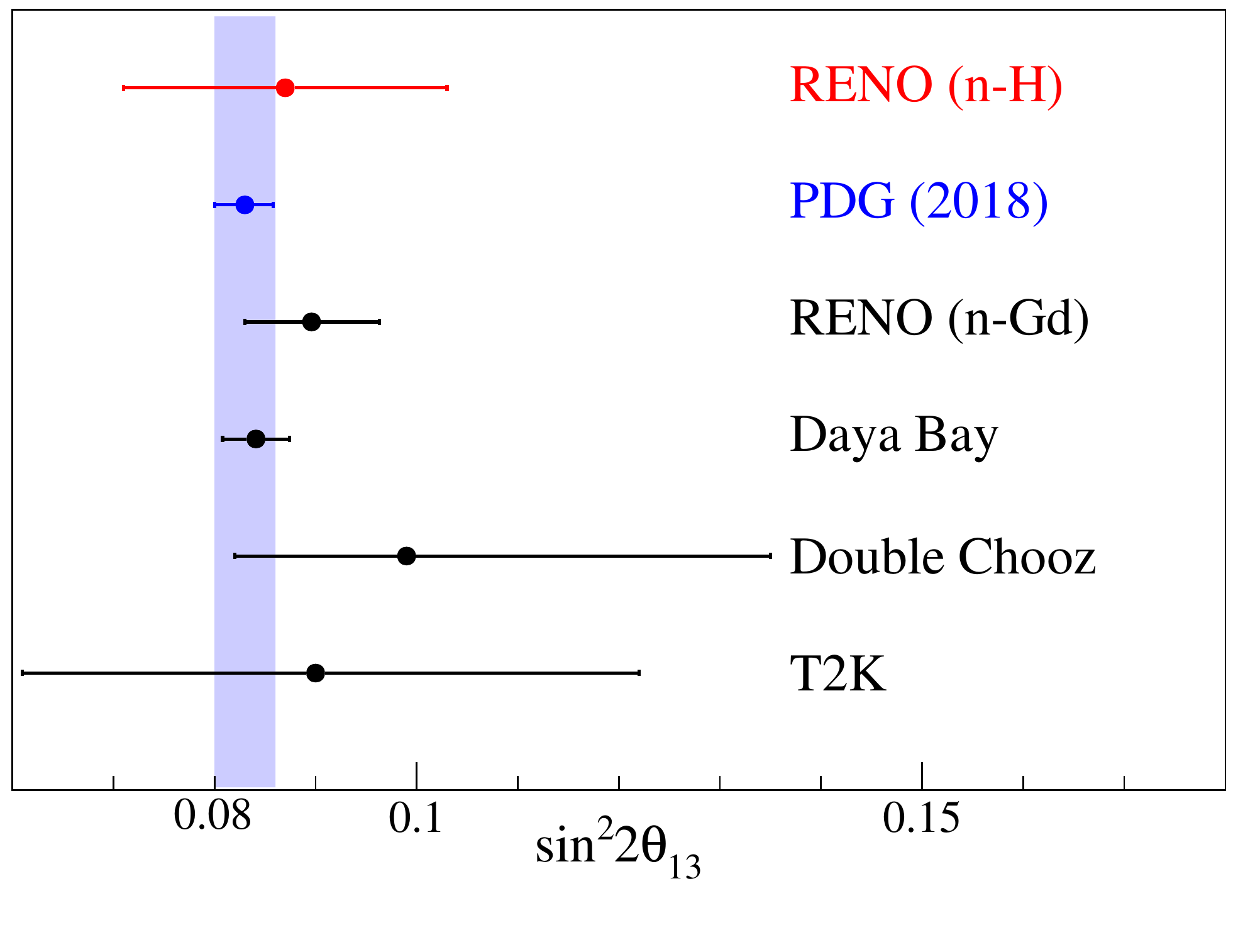}
	
	\caption{Comparison of current experimental results on sin$^{2}$2$\theta _{13}$. The blue shaded band represents the world average value from the Ref. \cite{PDG2018}. This result of sin$^{2}$2$\theta _{13}$ is consistent with those of RENO (n-Gd) \cite{RENOPRL3}, Daya Bay \cite{DayaBay2017}, Double Chooz \cite{DC2014}, and T2K \cite{T2K2017}.}
		\label{fig:CompWorld}
\end{figure}


In summary, RENO has performed an independent measurement of sin$^{2}$2$\theta _{13}$ via neutron capture on hydrogen using 1500 live days data, providing a result consistent with that of the n-Gd analysis. 
The measured value is compared with those of Daya Bay and Double Chooz experiments and found to be consistent within their errors as shown in Fig. \ref{fig:CompWorld}. In addition, if this result is combined with the n-Gd result, the uncertainty of the current $\theta _{13}$ measurement can be reduced by about 20\%. 
The error of sin$^{2}$2$\theta _{13}$ comes mostly from the systematic uncertainties of the backgrounds, detection efficiency and reactor. Future improvement of the systematic uncertainties will allow determination of both oscillation amplitude and frequency by a spectral analysis even using the n-H data sample. More precise measurements of sin$^{2}$2$\theta _{13}$ are necessary for constraining the leptonic CP phase if combined with the experimental results using accelerator neutrino beams. Independent IBD n-H measurements would provide additional information on the precise value of sin$^{2}$2$\theta _{13}$ as well as cross check.

\par
\vspace{5mm}

\section{Acknowledgments}

The RENO experiment is supported by the National Research Foundation of Korea (NRF) grant No. 2009-0083526, 2019R1A2C3004956, 2019R1A2B5B01070451, 2019R1I1A1A01059
548, 2016R1D1A3B02010606 
funded by the Korea Ministry of Science, ICT \& Future Planning. Some of us have been supported by a fund from the BK21 of NRF and Institute for Basic Science grant No. IBS-R017-G1-2019-a00. We gratefully acknowledge the cooperation of the Hanbit Nuclear Power Site and the Korea Hydro \& Nuclear Power Co. Ltd. (KHNP). We thank KISTI for providing computing and network resources through GSDC, and all the technical and administrative people who greatly helped in making this experiment possible.


\bibliographystyle{ieeetr}

\end{document}